\pdfoutput=1
\documentclass[11pt]{article}
\usepackage{defs}

\title{Solving One of~Many Instances Is~Almost as~Hard as~Solving One}
\author{
   Nikolai Chukhin
   \thanks{JetBrains Research.
   Email: \url{buyolitsez1951@gmail.com}}
   \and
   Alexander S. Kulikov
   \thanks{JetBrains Research.
   Email: \url{alexander.s.kulikov@gmail.com}}
   \and
   Ivan Mihajlin
   \thanks{JetBrains Research.
   Email: \url{ivmihajlin@gmail.com}}
   \and
   Alexander V. Smal
   \thanks{JetBrains Research.
   Email: \url{avsmal@gmail.com}}
}
\date{}

\begin{document}

\hypersetup{pageanchor=false}
\begin{titlepage}
\maketitle
\vspace{-2.5em}
\begingroup
\let\small\normalsize
\begin{abstract}
In~the 1-out-of-$k$ problem $\choice^k_F$ for a~function $F\colon A\times B\to\bits$, Alice gets $k$ distinct inputs $x_1,\ldots,x_k\in A$, Bob gets $k$ distinct inputs $y_1,\ldots,y_k\in B$, and they must output some~index $i\in\{1,\dotsc, k\}$ together with $F(x_i,y_i)$.
Clearly, the players can always compute $F(x_1,y_1)$, so the question is~whether the freedom to~choose the instance they solve can make the~problem substantially easier.
This question has been studied in~communication complexity and in~formula complexity, where it~is~related to~the KRW conjecture of~Karchmer, Raz, and Wigderson (1995), which would imply $\mathsf P\not\subseteq\mathsf{NC}^1$.

We~show that choosing among polynomially many instances saves at~most logarithmically many bits, precisely: for every Boolean function $F$ on~$n$-bit inputs and every $k\le2^n$,
\[\CC(\choice^k_F)\ge\CC(F)-O(\log k+\log n),\]
and the analogous bound holds for formula depth, which proves the conjectures of~Dinur and Meir (2016).
Previously, only a~square-root lower bound was known (Feder, Kushilevitz, Naor, and Nisan, 1995).

For the composition $U_m\diamond\KW_g$ of~the universal relation with the KW~game of~$g\colon\bits^n\to\bits$, we~prove that every~$g$ has a~core $S\subseteq\bits^n$ such that the game restricted to~any half of~$S^m$ still requires $\FD(g)-O(\log n)$ bits, where $\FD(g)$ is~the formula depth of~$g$.
For almost every~$g$, this holds even for $S=\bits^n$ and any $2^{1-m}$ fraction of~the inputs when $m\le n^2$, which implies
\[\CC(U_m\diamond\KW_g)\ge m+n-O(\log n),\]
improving the bound $m+n-O(\sqrt m)$ of~Wu (2023).
 \end{abstract}
\endgroup
\thispagestyle{empty}
\end{titlepage}
\hypersetup{pageanchor=true}
\setcounter{page}{1}

\section{Solving One of~Many Instances}\label{sec:introduction}

Suppose that Alice and Bob must compute a~function $F\colon A\times B\to\bits$, but instead of~one input pair they receive $k$ pairs $(x_1,y_1),\ldots,(x_k,y_k)$, with pairwise-distinct inputs on~each side, and must output $(i,F(x_i,y_i))$ for some $i\in[k]$.
We~denote this problem by~$\choice^k_F$.
Answering the first pair shows that this problem is~never harder than computing~$F$, and we~ask whether it~can be~substantially easier.
It~should be~noted that the distinctness condition is~essential: if~repeated inputs were permitted, then providing each player with $k$ copies of~a~single input would demonstrate that the problem is~at~least as~hard as~$F$.

Variants of~this question go~back to~selection problems in~computability and complexity theory~\cite{Jockusch68,Selman79,Ko83,BKS95} and in communication complexity~\cite{ABGKT01,CDKLM17}.
The closest to~our problem is~the choice problem $\choice_{F_1,\ldots,F_k}$ of~Beimel, Ben Daniel, Kushilevitz, and Weinreb~\cite{BBKW10,BBKW14}, in~which the players receive instances of~$k$ possibly different functions $F_1,\ldots,F_k$ and must output $(i,F_i(x_i,y_i))$ for some~$i$.
Since the functions may differ, repeated inputs do~not make this problem trivial, and they are allowed, so $\choice^k_F$ is~the restriction of~$\choice_{F,\ldots,F}$ to~distinct inputs.
Their motivation was the direct-sum question, which asks whether computing all $k$ values $F_1(x_1,y_1),\ldots,F_k(x_k,y_k)$ requires about $\sum_{j=1}^k\CC(F_j)$ bits, that~is, about as~many as~solving the instances separately~\cite{FKNN95,KRW95}.
Indeed, the players can compute all $k$ values by~repeatedly solving the choice problem on~the remaining instances and removing the solved one, hence even a~weak direct-sum theorem $\CC(F^k)\ge c\cdot k\cdot\CC(F)$ for distinct inputs would imply $\CC(\choice^k_F)\ge(c/2)\CC(F)$.

The direct-sum question itself was studied extensively in~different models of~computation.
The direct sum is~known to~hold for decision trees~\cite{JKS10} and for formula size~\cite{Wegener87}, and up~to~an~additive $O(k\log n)$ term for nondeterministic communication~\cite{KKN95}.
It~fails for Boolean circuits, since by~a~theorem of~Uhlig~\cite{Uhlig74,Wegener87}, many copies of~a~function of~near-maximal complexity can be~computed by~a~circuit of~about the size of~one copy, and for tensor rank, Shitov~\cite{Shitov19} refuted Strassen's direct-sum conjecture.
For randomized communication, the exact direct sum fails already for equality with private coins~\cite{FKNN95}, the distributional direct sum fails~\cite{BR14,GKR16}, and the general lower bound of~Barak, Braverman, Chen, and Rao loses a~factor of~about~$\sqrt k$~\cite{BBCR13}.
For deterministic communication, the strong form of~direct sum fails~\cite{MS25}, and whether a~weaker form $\CC(F^k)=\Omega(k\cdot\CC(F))$ holds remains open.
For the universal relation, in~which the players must find a~coordinate where two different strings differ, the direct sum holds almost exactly: a~rank argument of~Kozachinsky shows that $k$ instances on~$n$ bits require $k(n-1)-1$ bits~\cite[Remark~6]{Meir18}.

One motivation for both the direct-sum and the 1-out-of-$k$ questions comes from formula lower bounds.
Proving that some function in~$\mathsf P$ requires De~Morgan formulas of~depth $\omega(\log n)$, or~equivalently of~superpolynomial size, would show that $\mathsf P\not\subseteq\mathsf{NC}^1$, yet the best known size lower bounds for explicit functions are only about~$n^3$~\cite{Hastad98,Tal14}.
Karchmer, Raz, and Wigderson~\cite{KRW95} proposed to~approach this separation through composition of~Boolean functions.
For $f\colon\bits^m\to\bits$ and $g\colon\bits^n\to\bits$, the function $f\diamond g\colon\bits^{m\times n}\to\bits$ applies $g$ to~each row of~its input matrix and then $f$ to~the $m$ resulting bits, and the Karchmer--Raz--Wigderson (KRW) conjecture asserts that the formula depth of~$f\diamond g$ is~about the sum of~the formula depths of~$f$ and~$g$.
Iterating such compositions would give a~function in~$\mathsf P$ that requires depth $\omega(\log n)$, and hence $\mathsf P\not\subseteq\mathsf{NC}^1$.
Karchmer and Wigderson~\cite{KW90} showed that the formula depth of~$f$ equals the communication complexity of~the game $\KW_f$, in~which Alice receives $x\in f^{-1}(0)$, Bob receives $y\in f^{-1}(1)$, and they must find a~coordinate where $x$ and $y$ differ.
In~these terms, the KRW conjecture says that the natural protocol for $\KW_{f\diamond g}$ is~essentially optimal: the players first find a~row~$i$ with $g(X_i)\ne g(Y_i)$ by~solving $\KW_f$ on~the vectors of~row values, and then solve $\KW_g$ on~this row.
Karchmer, Raz, and Wigderson related this conjecture to~the direct-sum question: the $m$ rows are $m$ instances of~$\KW_g$, and the conjecture says that working on~all of~them at~once does not help.
The difficulty is~that a~protocol need not commit to~a~row in~advance: it~may solve $\KW_g$ on~whichever row with $g(X_i)\ne g(Y_i)$ turns out to~be~convenient.
The 1-out-of-$k$ problem captures this freedom: if~Alice gets $k$ distinct inputs in~$g^{-1}(0)$ and Bob gets $k$ distinct inputs in~$g^{-1}(1)$, then every pair is~a~valid instance of~$\KW_g$ and the players may answer any of~them, so~a~composition lower bound of~this kind requires that choosing does not make $\KW_g$ easier.

Dinur and Meir~\cite{DM16} formulated the problem with pairwise-distinct inputs in~their program toward the KRW conjecture.
They conjectured that choosing an~instance saves at~most polylogarithmically many bits, both in~communication complexity and in~formula depth~\cite[Conjectures~9.6 and~9.7]{DM16}.
Yet for general~$F$ the best known bound was only $\CC(F)\le O\bigl((\CC(\choice^k_F)+\log n)^2\bigr)$, which follows from bounds of~Feder, Kushilevitz, Naor, and Nisan~\cite{FKNN95}, as~observed by~Dinur and Meir~\cite{DM16}, and stronger bounds were known only in~special cases, such as~random functions.
For a~random function $f\colon\bits^n\to\bits$ and up~to~$\varepsilon2^n$ distinct instances, Meir and Tal~\cite{MT18} proved that choice and agreement are almost as~hard as~$f$ for circuits, formulas, and decision trees.
In~randomized query complexity, Bassilakis, Drucker, G\"o\"os, Hu, Ma, and Tan~\cite{BDGHMT20} proved that, for every function, analogues of~agreement and choice on~independent samples from a~suitable distribution require a~constant fraction of~its complexity.
We~prove that $\CC(\choice^k_F)\ge\CC(F)-O(\log k+\log n)$, hence resolving the conjectures of~Dinur and Meir for all $k$ up~to~$2^{\polylog(n)}$.

\subsection{Choosing Is~Almost as~Hard as~Computing}

\paragraph{Deterministic communication complexity.}
For a~total Boolean function $F\colon A\times B\to\bits$, the \emph{1-out-of-$k$ problem} $\choice^k_F$ gives Alice pairwise-distinct inputs $x_1,\ldots,x_k\in A$ and Bob pairwise-distinct inputs $y_1,\ldots,y_k\in B$.
They must output an~index $i\in[k]$ and the value $F(x_i,y_i)$.

\begin{restatable}{theorem}{mainBooleanChoice}\label{thm:global-boolean}
	For every total $F\colon A\times B\to\bits$ with $|A|,|B|\le2^n$, and every $1\le k\le\min\{|A|,|B|\}$,
	\[\CC(F)\le\CC(\choice^k_F)+O(\log k+\log n).\]
\end{restatable}

Previously, the best bound for general~$F$ was $\CC(F)\le O\bigl((\CC(\choice^k_F)+\log n)^2\bigr)$~\cite{FKNN95}.
\Cref{thm:global-boolean} replaces this quadratic bound by~an~additive loss of~$O(\log n)$ for polynomial~$k$ and of~$\polylog(n)$ for all $k\le2^{\polylog(n)}$, resolving the conjecture of~\cite{DM16}.

We~obtain \Cref{thm:global-boolean} from a~more general theorem for total \emph{search} problems.
Let $\mathcal R\subseteq A\times B\times O$ be~a~total relation, so~every pair $(x,y)\in A\times B$ has at~least one valid output $o\in O$.
In~$\choice^k_{\mathcal R}$, each player receives $k$ pairwise-distinct inputs and the players must find an~index $i\in[k]$ and an~output $o\in O$ with $(x_i,y_i,o)\in\mathcal R$.

\begin{restatable}{theorem}{mainRelationReconstruction}\label{thm:global-choice}
	Let $\mathcal R\subseteq A\times B\times O$ be~a~total relation, with $1 \le |A|,|B|\le2^n$.
	For every $1\le k\le\min\{|A|,|B|\}$,
	\[\CC(\mathcal R)\le\max\left\{\CC(\choice^k_{\mathcal R}),\lceil\log_2|O|\rceil\right\} +O(\log k+\log n).\]
\end{restatable}

\paragraph{Formula depth.}
For a~partial function $f\colon\bits^n\to\{0,1,*\}$, let $\FD(f)$ and $\FL(f)$ denote the minimum depth and leaf count of~a~fan-in-two De~Morgan formula computing~$f$. 
As~recalled above, $\FD(f)=\CC(\KW_f)$~\cite{KW90}.

The agreement function $\agree^k_f$ takes $k$ distinct inputs promised to~lie all in~$f^{-1}(0)$ or~all in~$f^{-1}(1)$ and returns their common value under~$f$.
It~is~the analogue of~choice for formulas: the KW~game of~$\agree^k_f$ is~$\choice^k_{\KW_f}$, so~$\FD(\agree^k_f)=\CC(\choice^k_{\KW_f})$.
Applying a~formula for~$f$ to~the first input shows that $\FD(\agree^k_f)\le\FD(f)$.
Agreement has also been studied before in \cite{BBKW14,DM16,MT18,Wu23}, but to~our knowledge no~nontrivial lower bound on~$\FD(\agree^k_f)$ was known for an~arbitrary~$f$, even for $k=2$.
Since $\FD(\agree^k_f)=\CC(\choice^k_{\KW_f})$, \Cref{thm:global-choice} with $O=[n]$ immediately gives the following bound.

\begin{restatable}{corollary}{mainAgreement}\label{thm:global-agreement}
	For every $k\ge1$ and every partial $f\colon\bits^n\to\{0,1,*\}$ with $|f^{-1}(0)|,|f^{-1}(1)|\ge k$,
	\[\FD(f)\le\FD(\agree^k_f)+O(\log k+\log n).\]
\end{restatable}

This proves the formula-depth conjecture of~Dinur and Meir~\cite{DM16} for any $k\le 2^{\polylog(n)}$, even for partial functions.

\paragraph{Exponentially many instances.}
For exponentially large~$k$, the additive $O(\log k)$ loss in~\Cref{thm:global-boolean,thm:global-agreement} can exceed the complexity of~a~single instance.
Nevertheless, \Cref{thm:global-choice} implies a~linear lower bound for every~$k$.

\begin{restatable}{corollary}{mainAllArity}\label{cor:global-all-arity}
	For $F,k$ as~in~\Cref{thm:global-boolean} and $f,k$ as~in~\Cref{thm:global-agreement},
	\begin{align*}
	 \CC(\choice^k_F)&\ge \Omega(\CC(F))-O(\log n),\\
	 \FD(\agree^k_f)&\ge \Omega(\FD(f))-O(\log n).
	\end{align*}
\end{restatable}

This follows because a~depth-$d$ protocol has at~most $2^d$ leaves and therefore can have at~most $2^d$ coordinates of~its input tuples as outputs, so~\Cref{thm:global-choice} applies with $2^d$ in~place of~$k$.
Once $\CC(F)$ exceeds a~sufficiently large multiple of~$\log n$, this improves the previous $\Omega(\sqrt{\CC(F)})-O(\log n)$ bound to~a~linear bound for every~$k$.

\paragraph{Two instances.}
For $k=2$, \Cref{thm:global-agreement} already gives an~$O(\log n)$ depth loss.
We~also give a~short alternative proof for this case by~adapting a~covering argument of~Beimel, Ben Daniel, Kushilevitz, and Weinreb~\cite{BBKW10} to~formulas; it~gives an~explicit additive loss in~depth and also bounds the formula size.

\begin{restatable}{theorem}{mainBinaryAgreement}\label{thm:binary-agreement}
	For every partial $f\colon\bits^n\to\{0,1,*\}$ with $|f^{-1}(0)|,|f^{-1}(1)|\ge2$,
	\begin{align*}
		\FD(f)&\le\FD(\agree^2_f)+\lceil\log_2 n\rceil+1,\\
		\FL(f)&\le2n\cdot\FL(\agree^2_f).
	\end{align*}
\end{restatable}

Note that these losses are necessary up~to~constants.
For $n\ge3$, the function $f_n(x)=\neg x_1\wedge(x_2\vee\cdots\vee x_n)$ depends on~all $n$ variables, so~it~needs at~least $n$ leaves and depth $\lceil\log_2 n\rceil$.
Yet $\agree^2_{f_n}$ is~computed by~$\neg x^{(1)}_1\wedge\neg x^{(2)}_1$, with two leaves and depth one.
Thus, for two instances, the formula depth can decrease by~at~most $\log n+O(1)$ and the formula size by~at~most a~factor of~$O(n)$, and both bounds are attained up~to~constants.

\subsection{Universal Composition and Dense Restrictions}

The freedom to~choose a~row is~only one way in~which a~protocol for $\KW_{f\diamond g}$ could beat the natural protocol.
More generally, every bit of~communication tells the players something about each other's inputs: after $t$ bits, following the larger half at~each step, they reach a~node whose rectangle contains at~least a~$2^{-t}$ fraction of~the inputs, and the rest of~the protocol must work only on~this rectangle.
To~show that the players still have to~solve $\KW_g$ there, one needs a~communication problem to~remain hard when its inputs are restricted to~a~dense subset of~the domain.
This dense-restriction question generalizes the 1-out-of-$k$ problem, which restricts the rows of~Alice's matrix to~$g^{-1}(0)$ and the rows of~Bob's matrix to~$g^{-1}(1)$.

A~standard approach toward the KRW conjecture replaces $\KW_f$ or~$\KW_g$ by~the universal relation $U_m$ (or~$U_n$), in~which the players receive two strings that are promised to~be~different and must find a~coordinate where they differ.
Analogues of~the KRW conjecture are known for $U_m\diamond U_n$~\cite{HW93,EIRS01} and for $\KW_f\diamond U_n$~\cite{GMWW17,KM18}, as~well as~when the inner function is~parity~\cite{DM16}, for classes of~monotone functions~\cite{dRMNPR24}, and for stronger composition relations~\cite{Meir25,CKM25}.
Here we~study the composition $U_m\diamond\KW_g$ of~the outer universal relation $U_m$ with the KW~game of~a~Boolean inner function~$g$, also considered by~Gavinsky, Meir, Weinstein, and Wigderson~\cite{GMWW17}.

In~$U_m\diamond\KW_g$, Alice and Bob receive matrices $X,Y\in\bits^{m\times n}$ and must output a~pair $(i,j)$ with $X_{i,j}\ne Y_{i,j}$, or~reject when their vectors of~row values $g^m(X)=(g(X_1),\ldots,g(X_m))$ and $g^m(Y)$ are equal.
Note that Alice can send $g^m(X)$ using~$m$ bits, after which the players can either reject or~solve $\KW_g$ on~a~row where $g^m(X)$ and $g^m(Y)$ differ, using $m+\FD(g)+O(1)$ bits in~total.
The goal is~to~prove a~matching lower bound of~roughly $m+\FD(g)$.

Mihajlin and Smal~\cite{MS21} showed that there is~an~inner function~$g\colon\bits^n\to\bits$ with $\CC(U_n\diamond\KW_g)\ge1.5n-o(n)$.
Wu~\cite{Wu23} improved it by showing that for almost every $g\colon\bits^n\to\bits$, when $\log^2 n=o(m)$ and $m=o(n^2)$,
\[\CC(U_m\diamond\KW_g)\ge m+n-O(\sqrt m).\]
He~asked whether the loss could be~made polylogarithmic.
We~reduce it~to~$O(\log n)$, for all $m\le n^2$.

\begin{restatable}{theorem}{mainUniversalComposition}\label{cor:random-universal-composition}
	With probability $1-o(1)$ as~$n\to\infty$, a~uniformly random $g\colon\bits^n\to\bits$ satisfies, for all $m\le n^2$,
	\[\CC(U_m\diamond\KW_g)\ge m+n-O(\log n) \ge m+\FD(g)-O(\log n).\]
\end{restatable}

Unlike~\cite{MS21,Wu23}, who reduce the problem that remains after about $m$ bits of~communication to~a~nondeterministic lower bound for non-equality, we~follow the dense-restriction approach of~Dinur and Meir~\cite{DM16}.
For a~subset $\mathcal X\subseteq\bits^{m\times n}$, we~write $\restr{g}{\mathcal X}$ for the restriction of~$U_m\diamond\KW_g$ to~$\mathcal X\times\mathcal X$.
Following a~protocol for $U_m\diamond\KW_g$ on~diagonal inputs $(X,X)$ for $m-1$ steps we can reach a~node whose rectangle contains $\mathcal X\times\mathcal X$ for a~subset $\mathcal X\subseteq\bits^{m\times n}$ of~density at~least $2^{1-m}$, where the remaining subprotocol must solve $\restr{g}{\mathcal X}$.
To~prove \Cref{cor:random-universal-composition}, it~therefore suffices to~show that $\restr{g}{\mathcal X}$ remains hard on~every such subset~$\mathcal X$, which we~do for almost every~$g$ in~\Cref{thm:random-dense-core}.

\paragraph{Arbitrary inner functions.}
\Cref{cor:random-universal-composition} holds only for almost every~$g$, while the KRW conjecture concerns every inner function.
We~therefore also study the dense-restriction approach for an~arbitrary~$g$: we~show what it~gives at~high densities and that it~cannot work uniformly at~the low densities that the composition requires.
An~arbitrary function~$g$ need not remain hard on~dense subsets of~the full domain $\bits^{m\times n}$, because $g$ may be~hard only on~a~small part of~$\bits^n$; for this reason, the question is~whether every $g$ has a~hard \emph{core} $\mathcal X_0\subseteq\bits^{m\times n}$ such that $\restr{g}{\mathcal X}$ remains hard for every sufficiently dense subset $\mathcal X\subseteq\mathcal X_0$~\cite[Conjecture~9.4]{DM16}.
As~a~first step, they asked whether $\CC(\restr{g}{\mathcal X})\ge\FD(g)-0.99m$ for every subset $\mathcal X\subseteq\mathcal X_0$ of~relative density~$1/2$ \cite[p.~3:48]{DM16}.

We~prove that every nonconstant function~$g$ has such a~core with only an~$O(\log n)$ loss.

\begin{restatable}{theorem}{mainHalfDensity}\label{thm:half-density-core}
	For every nonconstant $g\colon\bits^n\to\bits$, there is~a~nonempty $S\subseteq\bits^n$ such that, for every $m\ge2$ and every $\mathcal X\subseteq S^m$ of~relative density at~least $(1+2^{1-m})/3$,
	\[\CC(\restr{g}{\mathcal X})\ge \FD(g)-O(\log n).\]
\end{restatable}

We~note that the $O(\log n)$ loss is~tight up~to~a~constant factor (see~\Cref{sec:dense-negative-proof}).
In~particular, the answer to~the question of~Dinur and Meir depends on~$m$: their bound holds for every~$g$ when $m$ is~at~least a~large constant times $\log n$, but for $m\le(\log_2n)/2$ and all sufficiently large~$n$, there is~a~function~$g$ for which it~fails in~every core.

Moreover, Dinur and Meir~\cite[Conjecture~9.4]{DM16} conjectured that the approach also works at~the low densities that the composition requires: every~$g$ has a~core~$\mathcal X_0$ such that $\restr{g}{\mathcal X}$ loses at~most $\widetilde O(\sqrt m)$ bits for every $\mathcal X\subseteq\mathcal X_0$ of~relative density $2^{-m+\widetilde O(\sqrt m)}$, which, by~the argument above, would give $\CC(U_m\diamond\KW_g)\ge m+\FD(g)-\widetilde O(\sqrt m)$ for every~$g$.
We~show that this fails when both $\widetilde O(\sqrt m)$ terms depend only on~$m$: no~such bound holds uniformly over~$n$, even for an~arbitrary core $\mathcal X_0\subseteq\bits^{m\times n}$.

\begin{restatable}{theorem}{mainDenseCounterexample}\label{thm:uniform-density-counterexample}
	For every sufficiently large even $m$, there is~a~balanced Boolean function $g_m$ on~$n=2^{\Theta(m)}$ bits such that, for every nonempty core $\mathcal X_0\subseteq\bits^{m\times n}$ and every integer $0\le\sigma\le m-1$, some $\mathcal X\subseteq \mathcal X_0$ satisfies $|\mathcal X|\ge2^{-m+\sigma}|\mathcal X_0|$, but
	\[\CC(\restr{g_m}{\mathcal X})\le \FD(g_m)-m/2+\sigma+O(\log m).\]
\end{restatable}

Thus no~dense-core bound for subsets of~relative density $2^{-m+\sigma}$ with $\sigma=o(m)$ and with depth loss $o(m)$, where both $o(m)$ terms depend only on~$m$, can hold for all~$n$ and~$g$.
This refutes the uniform formulation of~\cite[Conjecture~9.4]{DM16}.
In~our construction, $n=2^{\Theta(m)}$, or~equivalently $m=\Theta(\log n)$, so~allowing an~additional $O(\log n)$ depth loss or~restricting to~$n=m^{O(1)}$ remains open (see~\Cref{sec:scope}).
 \section{Proof Ideas}\label{sec:proof-ideas}

In~this section, we~describe the main ideas of~our proofs.

\subsection{1-out-of-2 Agreement}\label{sec:overview-two}

For $k=2$, we~adapt the covering argument of~Beimel, Ben Daniel, Kushilevitz, and Weinreb~\cite{BBKW10}.
Let $H(u,v)$ be~a~formula for $\agree^2_f$, and consider its values on~the pairs in~$f^{-1}(0)\times f^{-1}(1)$.
A~greedy halving argument on~this matrix, finds at~most $n$ inputs $v^{(1)},\ldots,v^{(t)}\in f^{-1}(1)$ such that every $x\in f^{-1}(0)$ has $H(x,v^{(j)})=0$ for some~$j$, or~a~symmetric family in~$f^{-1}(0)$.
In~the first case,
\[f(x)=\bigwedge_{j=1}^t\left(H(x,v^{(j)})\vee\operatorname{EQ}_{v^{(j)}}(x)\right)\]
on~the promise: for $x\in f^{-1}(0)$, some term is~zero, and for $x\in f^{-1}(1)$, each term is~one by~the correctness of~$H$, except at~$x=v^{(j)}$, where the two inputs of~$H$ coincide and the equality test corrects the term.
This formula has depth $\FD(\agree^2_f)+O(\log n)$.
This proves \Cref{thm:binary-agreement}, see~\Cref{sec:agreement} for more details.

\subsection{The 1-out-of-\texorpdfstring{$k$}{k} Problem}\label{sec:overview-choice}

A~natural first attempt is~to~reduce the problem to~two instances.
Let $k$ be~a~power of~two, and split the $k$ instances into two halves.
Each half is~an~instance of~$\choice^{k/2}_{\mathcal R}$, and a~protocol for $\choice^k_{\mathcal R}$ solves one of~these two instances, so~it~is~a~protocol for $\choice^2_{\mathcal S}$ with $\mathcal S=\choice^{k/2}_{\mathcal R}$, restricted to~halves whose entries are all distinct.
Suppose that every relation $\mathcal S$ on~$N$-bit inputs satisfied $\CC(\mathcal S)\le\CC(\choice^2_{\mathcal S})+O(\log N)$, even on~this restriction.
Applying this bound to~$\choice^{2^i}_{\mathcal R}$, whose inputs have $2^in$ bits, for $i=\log_2k-1,\ldots,0$ would give
\[\CC(\mathcal R)\le\CC(\choice^k_{\mathcal R})+\sum_{i=0}^{\log_2k-1}O(i+\log n)=\CC(\choice^k_{\mathcal R})+O(\log k\cdot(\log k+\log n)).\]
This loses a~factor of~$\log k$ compared with \Cref{thm:global-choice}; for instance, for $k=\operatorname{poly}(n)$, it~gives a~loss of~$O(\log^2n)$ instead of~$O(\log n)$.
We~therefore handle all $k$ instances at~once.

Let $\Pi$ be~an~optimal protocol for $\choice^k_{\mathcal R}$, and let $x\in A$ and $y\in B$ be~inputs for~$\mathcal R$.
The natural way to~use~$\Pi$ is~to~put $x$ and~$y$ at~one position of~two tuples whose other entries are fixed, and to~run~$\Pi$ on~these tuples.
This fails because $\Pi$ may answer at~one of~the fixed positions, and then its output says nothing about $(x,y)$.

Our approach replaces a~single run of~$\Pi$ with a~game on~its tree.
Every leaf $w$ of~$\Pi$ names a~position~$i_w$ and an~output, and this output is~valid for every pair of~inputs that can stand at~position~$i_w$ in~tuples reaching~$w$.
On~single inputs $x$ and~$y$, the players walk down the tree of~$\Pi$ and choose the children at~their own nodes freely: Alice steers the walk to~leaves at~which $x$ can stand at~the named position, and Bob steers it~to~leaves at~which $y$ can.
If~both succeed, the leaf they reach gives a~valid output for $(x,y)$.
This raises two questions: for which pairs $(x,y)$ do~such strategies exist, and how do~the players find out that their pair is~one of~them?

Call a~rectangle $A'\times B'$ solvable if~some protocol of~depth $\max\{\CC(\choice^k_{\mathcal R}),\lceil\log_2|O|\rceil\}$ solves $\mathcal R$ on~it.
We~show that for all distributions $\mu$ on~$A$ and $\nu$ on~$B$, some solvable rectangle satisfies
\[\mu(A')+\nu(B')\ge1+k^{-2}.\]
This inequality says that $A'\times B'$ is~slightly larger than the trivial solvable rectangles $A\times\varnothing$ and $\varnothing\times B$, which give exactly one.
If~$\mu$ or~$\nu$ gives some input probability at~least $k^{-2}$, such a~rectangle is~easy to~find: for instance, if~$\mu(x_0)\ge k^{-2}$ for some $x_0\in A$, then $\{x_0\}\times B$ is~solvable, since Bob can send a~valid output for $(x_0,y)$, and $\mu(\{x_0\})+\nu(B)\ge1+k^{-2}$.
Otherwise, random tuples $X\sim\mu^k$ and $Y\sim\nu^k$ have distinct entries with probability close to~one.
Running $\Pi$ on~$X$, Alice can steer the walk to~leaves at~which her own entry stands at~the named position, and on~$Y$, Bob can do~the same.
We~judge the leaves at~positions up~to~$j$ by~Alice's tuple and the other leaves by~Bob's tuple, and move $j$ from $0$ to~$k$.
By~Zermelo's theorem~\cite{Zermelo13}, in~every such game one of~the players can force the outcome, so~at~some position~$i$ the probability that Alice wins increases by~at~least~$k^{-2}$.
Fixing all entries except the $i$th ones gives the rectangle (\Cref{lem:global-weak-separation}).

The players find a~solvable rectangle that contains $(x,y)$ with $O(\log k+\log n)$ bits of~communication.
By~the minimax theorem, there is~a~random solvable rectangle that contains each input of~either player with probability at~least $1/2+k^{-2}/4$; probability $1/2$ is~trivial, by~choosing $A\times\varnothing$ or~$\varnothing\times B$.
By~minimax, it~suffices that for every distribution on~$A\sqcup B$, some solvable rectangle contains a~random input with this probability; if~the distribution puts noticeably more weight on~one player, a~trivial rectangle does, and otherwise the inequality above does.
Majority amplification turns this random rectangle into a~formula of~depth $O(\log k+\log n)$ whose leaves are solvable rectangles.
Then, the players walk down this formula to~a~rectangle that contains $(x,y)$, and then solve $\mathcal R$ on~it.

\subsection{Dense Restrictions and Composition}\label{sec:overview-dense}

Recall that in~$U_m\diamond\KW_g$ the players receive matrices $X,Y\in\bits^{m\times n}$ and must find a~cell where they differ or~reject when $g^m(X)=g^m(Y)$, and that $\restr{g}{\mathcal X}$ is~the restriction of~this problem to~$\mathcal X\times\mathcal X$, where $\mathcal X\subseteq\bits^{m\times n}$.
We~first show that lower bounds on~$\CC(\restr{g}{\mathcal X})$ for all dense sets~$\mathcal X$ give lower bounds on~$\CC(U_m\diamond\KW_g)$.
Take a~protocol for $U_m\diamond\KW_g$ and walk down its tree, at~each step moving to~a~child that keeps at~least half of~the remaining diagonal inputs $(X,X)$, on~which both players have the same matrix $X\in\bits^{m\times n}$.
Let $\mathcal X$ be~the set of~all matrices~$X$ such that $(X,X)$ reaches the current node after $m-1$ steps; its density is~at~least $2^{1-m}$.
Hence a~lower bound on~$\CC(\restr{g}{\mathcal X})$ for all sets~$\mathcal X$ of~density at~least $2^{1-m}$ adds to~the $m-1$ bits already sent (see \Cref{lem:dense-composition-transfer}).

To~prove such lower bounds, we~turn a~protocol for $\restr{g}{\mathcal X}$ into a~formula.
Consider a~leaf which rejects, i.e. all the inputs reaching such a~leaf form a~rectangle, and every pair $(X,Y)$ in~it~satisfies $g^m(X)=g^m(Y)$, so~all these matrices have the same value $c\in\bits^m$ of~$g^m$, and we~label the leaf with a~new variable~$p_c$.
This gives a~formula $\phi(X,p)$ in~the bits of~a~matrix~$X$ and the variables $p=(p_c)_{c\in\bits^m}$ whose depth is~at~most that of~the protocol, and we~claim that $\phi(X,p)=p_{g^m(X)}$ for every $X\in\mathcal X$ and every~$p$ (\Cref{lem:dense-protocol-encoding}).
Indeed, if~$\phi(X,p)=1$, then we~can walk from the root to~a~leaf through nodes of~value one, following Alice's moves on~$X$ at~her nodes and choosing a~child of~value one at~Bob's nodes.
This leaf can be~reached with Alice's input~$X$, so~it~is~not a~literal, since the literal at~a~leaf is~zero on~all Alice's inputs that reach this leaf; hence it~is~$p_{g^m(X)}$, and $p_{g^m(X)}=1$.
The case $\phi(X,p)=0$ is~symmetric, with Bob in~place of~Alice.
Thus, $\phi$ determines $g^m(X)$, that is, the values of~$g$ on~all rows of~$X$, for every $X\in\mathcal X$.

For an~arbitrary function $g\colon\bits^n\to\bits$ and a~nonempty set $S\subseteq\bits^n$, if~$\mathcal X\subseteq S^m$ has relative density at~least $(1+2^{1-m})/3$, then substituting suitable values for~$p$ and a~hybrid argument over the rows turn~$\phi$ into a~formula of~depth $\CC(\restr{g}{\mathcal X})+1$ that computes~$g$ on~at~least two thirds of~$S$ (\Cref{lem:dense-grouped-prediction}).
It~remains to~find a~set~$S$ on~which no~formula of~depth $\FD(g)-O(\log n)$ computes~$g$ on~two thirds of~the inputs (\Cref{lem:dense-uniform-hard-set}).
The construction of~such a~set resembles Nisan's proof of~Impagliazzo's hard-core lemma~\cite{Impagliazzo95}.

For a~random~$g$, a~fixed formula~$\phi$ defines a~map from matrices to~$\bits^m\cup\{\bot\}$ that does not depend on~$g$ and that equals $g^m$ on~$\mathcal X$, that is, on~at~least a~$2^{1-m}$ fraction of~all matrices.
For a~fixed map, this happens with very small probability (\Cref{lem:dense-predictor-tail}), so~we take a~union bound over all formulas of~depth $n-O(\log n)$.

Finally, to~prove \Cref{thm:uniform-density-counterexample}, we~construct a~balanced function~$g$ that is~hard for formulas but easy for random formulas in~the following sense: $g$ has depth about $3r$, whereas a~random formula of~depth about~$2r$ computes $g(x)$ correctly at~every input~$x$ with probability close to~one (\Cref{lem:dense-graph-predictors}).
In~every core $\mathcal X_0$, some formula~$\psi$ from this distribution is~correct on~all $m$ rows of~at~least half of~the matrices.
Keeping only the matrices with the most frequent values of~$g^m$, we~obtain a~large subset~$\mathcal X$ on~which Alice sends $g^m(X)$ with few bits and the players then solve $\KW_\psi$ on~a~row where the values differ (\Cref{lem:dense-core-trimming}).
 \section{Preliminaries}\label{sec:preliminaries}

A~relation $\mathcal R\subseteq A\times B\times O$ is~\emph{total} if~every pair $(x,y)\in A\times B$ has an~output $o\in O$ with $(x,y,o)\in\mathcal R$.
A~partial function $f\colon\bits^n\to\{0,1,*\}$ is~\emph{nonconstant} if~both $f^{-1}(0)$ and $f^{-1}(1)$ are nonempty.

\subsection{Formulas and KW Games}\label{sec:prelim-formulas}

A~De Morgan formula is~a~binary tree whose inner nodes are AND and OR gates and whose leaves are literals $x_j$ and $\neg x_j$ or~constants.
Its depth is~the maximum number of~gates on~a~path from the root to~a~leaf.
A~formula computes a~partial function $f\colon\bits^n\to\{0,1,*\}$ if~it~is~correct on~the promise $f^{-1}(0)\cup f^{-1}(1)$, and we~write $\FD(f)$ and $\FL(f)$ for the minimum depth and the minimum number of~leaves of~such a~formula.

For fixed $z\in\bits^n$, the equality test
\[\operatorname{EQ}_z(x)=\bigwedge_{j=1}^n[x_j=z_j]\]
has depth $\lceil\log_2n\rceil$ and exactly $n$ leaves.

Let $\mathcal R\subseteq A\times B\times O$ be~a~total relation.
A~deterministic protocol for $\mathcal{R}$ is~a~binary tree in~which every inner node belongs to~Alice or~to~Bob and every leaf is~labeled by~an~output $o\in O$.
Each node of~Alice is~labeled by~a~function $A\to\bits$, and each node of~Bob by~a~function $B\to\bits$, which tells which child to~go to.
On~inputs $x\in A$ and $y\in B$, the players start at~the root and repeatedly go to~the child chosen by~the owner of~the current node on~its own input, until they reach a~leaf, whose label is~the output.
The protocol solves $\mathcal R$ if~its output $o$ satisfies $(x,y,o)\in\mathcal R$ for all $x\in A$ and $y\in B$.
The depth of~a~protocol is~the maximum length of~a~path from the root to~a~leaf, that is, the number of~bits sent in~the worst case.
We~write $\CC(\mathcal R)$ for the minimum depth of~a~protocol that solves $\mathcal R$.

Karchmer and Wigderson~\cite{KW90} showed that formula depth can be~expressed as~the communication complexity of~a~game associated with the function.

\begin{definition}\label{def:kw}
	Let $f\colon\bits^n\to\{0,1,*\}$ be~a~nonconstant partial function.
	The Karchmer--Wigderson game of~$f$ is~the following relation:
	\[\KW_f=\{(x,y,j)\in f^{-1}(0)\times f^{-1}(1)\times[n]:x_j\ne y_j\}.\]
	Thus Alice gets $x\in f^{-1}(0)$, Bob gets $y\in f^{-1}(1)$, and they must output a~coordinate~$j$ with $x_j\ne y_j$.
\end{definition}

\begin{theorem}[{\cite{KW90}}]\label{thm:kw}
	For every nonconstant partial function~$f$, \[\FD(f)=\CC(\KW_f).\]
\end{theorem}

\subsection{Agreement and 1-out-of-\texorpdfstring{$k$}{k} Problems}\label{sec:prelim-agreement}

We~first define the agreement function, which asks for the common value of~a~partial function on~$k$ distinct inputs.
For a~set $A$ and $k\ge1$, let $\Inj_k(A)$ be~the set of~$k$-tuples of~pairwise distinct elements of~$A$:
\[\Inj_k(A)=\{(a_1,\ldots,a_k)\in A^k:a_i\ne a_j\text{ for }i\ne j\}.\]

\begin{definition}\label{def:agreement}
	Let $f\colon\bits^n\to\{0,1,*\}$ and let $k\ge1$.
	If~$|f^{-1}(0)|,|f^{-1}(1)|\ge k$, the agreement function $\agree^k_f$ is~the partial function on~$kn$ bits defined by
	\[(\agree^k_f)^{-1}(b)=\Inj_k(f^{-1}(b))\quad\text{for }b\in\bits,\]
	and $*$ elsewhere.
\end{definition}
That is, the input consists of~$k$ distinct strings $x^{(1)},\ldots,x^{(k)}\in\bits^n$ with the promise that $f(x^{(1)})=\dotsb=f(x^{(k)})\in\bits$, and the output is~this common value.

Now, we~define the 1-out-of-$k$ problem for every total relation.

\begin{definition}\label{def:distinct-oneout}
	Let $\mathcal R\subseteq A\times B\times O$ be~a~total relation.
	For $1\leq k\leq\min\{|A|,|B|\}$, the 1-out-of-$k$ problem $\choice^k_{\mathcal R}$ gives Alice $X=(x_1,\ldots,x_k)\in\Inj_k(A)$ and Bob $Y=(y_1,\ldots,y_k)\in\Inj_k(B)$ and asks for a~pair $(i,o) \in [k] \times O$ such that $(x_i,y_i,o)\in\mathcal R$.
\end{definition}

We~identify a~function $F\colon A\times B\to\bits$ with the total relation $\{(x,y,F(x,y)):x\in A,\,y\in B\}$.

\begin{remark}\label{rem:agreement-kw}
	If~$|f^{-1}(0)|,|f^{-1}(1)|\ge k$, the KW game of~$\agree^k_f$ is~$\choice^k_{\KW_f}$, where an~output $(i,j)$ names the $j$th bit of~the $i$th input.
	Indeed, Alice gets distinct $x_1,\ldots,x_k\in f^{-1}(0)$, Bob gets distinct $y_1,\ldots,y_k\in f^{-1}(1)$, and the $j$th bits of~$x_i$ and~$y_i$ differ exactly when $(x_i,y_i,j)\in\KW_f$.
	Hence, by~\Cref{thm:kw},
	\[\FD(\agree^k_f)=\CC(\choice^k_{\KW_f}).\]
\end{remark}

\subsection{Minimax and Majority Amplification}\label{sec:prelim-minimax}

We~use von Neumann's minimax theorem~\cite{vonNeumann28} for zero-sum games.

\begin{lemma}[\cite{vonNeumann28}]\label{lem:minimax}
	Let $\mathcal H$ and $\mathcal Z$ be~finite nonempty sets, let $C\subseteq\mathcal H\times\mathcal Z$, and let $\delta\in\mathbb R$.
	If~for every distribution~$\pi$ on~$\mathcal Z$ some $H\in\mathcal H$ satisfies $\Pr_{z\sim\pi}[(H,z)\in C]\ge\delta$, then some distribution~$\Lambda$ on~$\mathcal H$ satisfies $\Pr_{H\sim\Lambda}[(H,z)\in C]\ge\delta$ for every $z\in\mathcal Z$.
\end{lemma}

The lemma below turns a~distribution over formulas that is~correct with probability $1/2+\varepsilon$ at~every point into a~single formula that is~correct at~every point.

\begin{lemma}[\cite{Valiant84}]\label{lem:majority-amplification}
	Let $\mathcal Z$ be~a~finite set, let $f\colon\mathcal Z\to\bits$, let $\mathcal F$ be~a~set of~functions $\mathcal Z\to\bits$, and let $0<\varepsilon\le1/2$.
	If~a~distribution~$\Lambda$ over formulas of~depth at~most~$d$ with AND and OR gates and leaves from~$\mathcal F$ satisfies $\Pr_{H\sim\Lambda}[H(z)=f(z)]\ge1/2+\varepsilon$ for every $z\in\mathcal Z$, then some formula $H^*$ of~depth $d+O(\log(1/\varepsilon)+\log\log|\mathcal Z|)$ with leaves from~$\mathcal F$ satisfies $H^*(z)=f(z)$ for every $z\in\mathcal Z$.
\end{lemma}
 \section{A~Simple Proof for 1-out-of-2 Agreement}\label{sec:agreement}

In~this section, we~give a~simple proof of~\Cref{thm:binary-agreement}, which takes its main idea from the proof of~\cite[Theorem~4]{BBKW10}.

\mainBinaryAgreement*
\begin{proof}
	Let $H$ be~any formula for $\agree^2_f$ with depth $d$ and $\ell$ leaves.
	We~prove that 
	\[\FD(f)\le d+\lceil\log_2n\rceil+1,\qquad \FL(f)\le\min\{n,\ell\}(\ell+n).\]

	Write the input to~$H$ as~two binary strings $(u,v)$, where $u,v\in\bits^n$.
	The variable in~each leaf of~$H$ belongs either to~the first string~$u$ or~to~the second string~$v$.
	Let $n_u$ and $n_v$ be~the numbers of~distinct variables of~$u$ and of~$v$, respectively, that appear in~the leaves, so~$n_u+n_v\le\ell$.
	If~$n_u=0$ or~$n_v=0$, then $H$ depends on~only one string, and it~already computes $f$ on~that string.
	So, from now on, we~assume that $n_u,n_v\ge1$.

	Consider the Boolean matrix on~$F_0\times F_1$, where $F_0=f^{-1}(0)$ and $F_1=f^{-1}(1)$, with entries $H(u,v)$.
	Since $H$ reads at~most $n_u$ variables from~$u$ and at~most $n_v$ variables from~$v$, and since $|F_0|,|F_1|<2^n$, keeping one representative for each distinct row and column gives an~$r\times s$ matrix~$M$ with $r\le\min\{2^{n_u},2^n-1\}$ and $s\le\min\{2^{n_v},2^n-1\}$.
	We~claim that for some $t\le\min\{n,\ell\}$,
	\[\exists\,v^{(1)},\ldots,v^{(t)}\in F_1  \colon  \forall x\in F_0\ \exists j,  H(x,v^{(j)})=0,\]
	or
	\[\exists\,u^{(1)},\ldots,u^{(t)}\in F_0  \colon  \forall x\in F_1\ \exists j,  H(u^{(j)},x)=1.\]
	To~prove this, we~keep a~set~$R$ of~uncovered rows and a~set~$C$ of~uncovered columns of~$M$, initially all $r$ rows and all $s$ columns.
	If~at~least half of~the entries of~$M$ in~$R\times C$ are zeros, then some column in~$C$ has zeros in~at~least half of~the rows in~$R$, and we~pick it and remove these rows from~$R$.
	Otherwise, symmetrically, some row in~$R$ has ones in~more than half of~the columns in~$C$, and we~pick it and remove these columns from~$C$.
	Each picked column halves~$R$ and each picked row halves~$C$, so~before $R$ or~$C$ becomes empty, we~pick at~most $\lceil\log_2(r+1)\rceil$ columns and at~most $\lceil\log_2(s+1)\rceil$ rows.
	If~$R$ becomes empty, the picked columns give the first alternative, and if~$C$ becomes empty, the picked rows give the second.
	Since $n_u+n_v\le\ell$ with $n_u,n_v\ge1$, we~have $\lceil\log_2(r+1)\rceil\le\min\{n_u+1,n\}\le\min\{n,\ell\}$ and similarly $\lceil\log_2(s+1)\rceil\le\min\{n,\ell\}$, and hence $t \le \min \{ n, \ell \}$.

	In~the first case, set
	\[G(x)=\bigwedge_{j=1}^t\left(H(x,v^{(j)})\vee\operatorname{EQ}_{v^{(j)}}(x)\right).\]
	For $x\in F_0$, we~have $\operatorname{EQ}_{v^{(j)}}(x)=0$ for every~$j$ because $v^{(j)}\in F_1$, and $H(x,v^{(j)})=0$ for at~least one~$j$ by~the covering property, so~$G(x)=0$.
	For $x\in F_1$, each factor is~one: if~$x=v^{(j)}$, then $\operatorname{EQ}_{v^{(j)}}(x)=1$, and if~$x\ne v^{(j)}$, then $(x,v^{(j)})\in\Inj_2(F_1)$ and hence $H(x,v^{(j)})=1$.
	Thus $G$ computes $f$.
	The second case is~symmetric.

	Each term has depth at~most $\max\{d,\lceil\log_2n\rceil\}+1$ and at~most $\ell+n$ leaves, and a~balanced tree over the $t$ terms adds depth $\lceil\log_2t\rceil\le\min\{d,\lceil\log_2n\rceil\}$, since $t\le\min\{n,\ell\}$ and $\ell\le2^d$.
	Hence $G$ has depth at~most $\max\{d,\lceil\log_2n\rceil\}+\min\{d,\lceil\log_2n\rceil\}+1=d+\lceil\log_2n\rceil+1$ and at~most $t(\ell+n)\le\min\{n,\ell\} \cdot (\ell+n) \le 2 n \ell$ leaves.
\end{proof}
 \section{The 1-out-of-\texorpdfstring{$k$}{k} Problem}\label{sec:global-choice}

In this section, we prove our main result, \Cref{thm:global-choice}.

\mainRelationReconstruction*

Before proving \Cref{thm:global-choice}, we~derive the Boolean-function and agreement consequences stated in~the introduction.

\mainBooleanChoice*
\begin{proof}
	Let $D=\CC(\choice^k_F)$.
	By~\Cref{thm:global-choice} with the output set~$\bits$,
	\[\CC(F)\le\max\{D,1\}+O(\log k+\log n)\le D+O(\log k+\log n).\qedhere\]
\end{proof}

\mainAgreement*
\begin{proof}
	By~\Cref{thm:kw} and \Cref{rem:agreement-kw}, we~have $\FD(f)=\CC(\KW_f)$ and $\FD(\agree^k_f)=\CC(\choice^k_{\KW_f})$.
	The relation $\KW_f\subseteq f^{-1}(0)\times f^{-1}(1)\times[n]$ is~total because $f^{-1}(0)$ and $f^{-1}(1)$ are disjoint.
	Since moreover $|f^{-1}(0)|,|f^{-1}(1)|\le2^n$ and $k\le\min\{|f^{-1}(0)|,|f^{-1}(1)|\}$, \Cref{thm:global-choice} with the output set~$[n]$ gives
	\begin{align*}
		\FD(f)&\le\max\{\FD(\agree^k_f),\lceil\log_2n\rceil\}+O(\log k+\log n) \\
			  &\le\FD(\agree^k_f)+O(\log k+\log n).\qedhere
	\end{align*}
\end{proof}

The proof of~\Cref{thm:global-choice} combines protocols that are correct only on~part of~the inputs.
Throughout, $\mathcal R\subseteq A\times B\times O$ is~the total relation of~\Cref{thm:global-choice}: Alice's inputs lie in~$A$, Bob's inputs lie in~$B$, and $O$ is~the set of~outputs.
For $A'\subseteq A$ and $B'\subseteq B$, a~protocol \emph{solves $\mathcal R$ on~$A'\times B'$} if, whenever Alice receives $x\in A'$ and Bob receives $y\in B'$, the output~$o$ at~the leaf they reach satisfies $(x,y,o)\in\mathcal R$.
We~call the rectangle $A'\times B'$ \emph{solvable} if~some protocol of~depth at~most $\max\{\CC(\choice^k_{\mathcal R}),\lceil\log_2|O|\rceil\}$ solves $\mathcal R$ on~it.
Also, we~regard a~rectangle as~the pair of~its sides, so~that $A\times\varnothing$ and $\varnothing\times B$ are different solvable rectangles.

Fix an~optimal protocol~$\Pi$ for $\choice^k_{\mathcal R}$, of~depth $\CC(\choice^k_{\mathcal R})$.
We~follow the plan of~\Cref{sec:overview-choice}: \Cref{lem:global-weak-separation} below gives a~large solvable rectangle for each pair of~distributions, and the proof of~\Cref{thm:global-choice} combines these rectangles by~minimax and majority into a~protocol for~$\mathcal R$.

For each node~$v$ of~$\Pi$, let $\mathcal A_v\subseteq\Inj_k(A)$ be~the set of~tuples on~which Alice's messages agree with the edges leaving her nodes on~the path from the root to~$v$, and define $\mathcal B_v\subseteq\Inj_k(B)$ in~the same way for Bob.
For each leaf~$w$ with output $(i_w,o_w)$, let $S_w=\{x_{i_w}:(x_1,\ldots,x_k)\in\mathcal A_w\}$ and $T_w=\{y_{i_w}:(y_1,\ldots,y_k)\in\mathcal B_w\}$ be~the projections of~$\mathcal A_w$ and $\mathcal B_w$ onto coordinate~$i_w$.
Since every pair of~tuples in~$\mathcal A_w\times\mathcal B_w$ reaches~$w$ and receives the output $(i_w,o_w)$, we~have
\begin{equation}\label{eq:global-leaf-output}
	(x,y,o_w)\in\mathcal R\qquad\text{for all }x\in S_w\text{ and }y\in T_w.
\end{equation}

In~the end, Alice and Bob hold single inputs $x$ and~$y$ instead of~tuples, so~they cannot simply run~$\Pi$; instead, they walk down its tree and choose the children at~their nodes freely.
For a~set~$L$ of~leaves of~$\Pi$, Alice \emph{can force}~$L$ if~she can choose one child at~each of~her nodes so~that the walk ends in~$L$ whatever children Bob chooses at~his nodes; Bob can force~$L$ in~the symmetric sense.

\begin{lemma}[\cite{Zermelo13}]\label{lem:global-determinacy}
	For every set~$L$ of~leaves of~$\Pi$, either Alice can force~$L$ or~Bob can force the complement of~$L$, but not both.
\end{lemma}

The following lemma is~the main part of~our argument.

\begin{lemma}\label{lem:global-weak-separation}
	For every distribution~$\mu$ on~$A$ and distribution~$\nu$ on~$B$, some solvable rectangle $A'\times B'$ satisfies
	\[\mu(A')+\nu(B')\ge1+k^{-2}.\]
\end{lemma}
\begin{proof}
	Suppose first that $\mu(x_0)\ge k^{-2}$ for some $x_0\in A$.
	The rectangle $\{x_0\}\times B$ is~solvable: Bob uses $\lceil\log_2|O|\rceil$ bits to~announce some $o\in O$ with $(x_0,y,o)\in\mathcal R$, which exists by~totality, and the players output~$o$.
	Since $\mu(x_0)+\nu(B)\ge1+k^{-2}$, this rectangle satisfies the claim.
	Symmetrically, if~$\nu(y_0)\ge k^{-2}$ for some $y_0\in B$, then Alice announces an~output that is~valid for $(x,y_0)$, and the rectangle $A\times\{y_0\}$ satisfies the claim.

	Otherwise, draw $X=(x_1,\ldots,x_k)\sim\mu^k$ and $Y=(y_1,\ldots,y_k)\sim\nu^k$ independently.
	For any two distinct indices, the collision probability under~$\mu$ is~$\sum_{x\in A}\mu(x)^2\le\max_{x\in A}\mu(x)<k^{-2}$, so~a~union bound over the $\binom{k}{2}$ pairs gives $\Pr[X\notin\Inj_k(A)]\le\binom{k}{2}k^{-2}$, and similarly $\Pr[Y\notin\Inj_k(B)]\le\binom{k}{2}k^{-2}$.
	For $0\le j\le k$, let $p_j$ be~the probability that Alice can force the set
	\[L_j=\{w:i_w\le j,\ x_{i_w}\in S_w\}\cup\{w:i_w>j,\ y_{i_w}\notin T_w\},\]
	in~which leaves at~positions up~to~$j$ are judged by~Alice's tuple and the other leaves by~Bob's tuple.
	If~$X\in\Inj_k(A)$, Alice can force~$L_k$ by~following the messages of~$\Pi$ on~$X$, since every leaf~$w$ reached in~this way satisfies $X\in\mathcal A_w$ and hence $x_{i_w}\in S_w$.
	Similarly, if~$Y\in\Inj_k(B)$, Bob can force the complement of~$L_0$, and then Alice cannot force~$L_0$ by~\Cref{lem:global-determinacy}.
	Hence $p_k-p_0\ge1-2\binom k2k^{-2}=1/k$, so~some index $i\in[k]$ satisfies $p_i-p_{i-1}\ge k^{-2}$.
	By~averaging over all coordinates $x_j$ and $y_j$ with $j\ne i$, we~can fix their values so~that the difference $p_i-p_{i-1}$ remains at~least~$k^{-2}$; the coordinates $x_i\sim\mu$ and $y_i\sim\nu$ remain independent.

	After this fixing, $L_{i-1}$ and $L_i$ contain the same set~$L'$ of~leaves~$w$ with $i_w\ne i$; let $L''$ be~the set of~the remaining leaves~$w$ with $i_w\ne i$.
	Let $A'$ be~the set of~$x\in A$ for which Alice can force $L'\cup\{w:i_w=i,\ x\in S_w\}$, and let $B'$ be~the set of~$y\in B$ for which Bob can force $L''\cup\{w:i_w=i,\ y\in T_w\}$.
	The rectangle $A'\times B'$ is~solvable: on~inputs $x\in A'$ and $y\in B'$, the players walk down the tree of~$\Pi$, which has depth $\CC(\choice^k_{\mathcal R})$; at~each node, its owner sends one bit naming the child prescribed by~a~forcing strategy for the set in~the definition of~$A'$ or~$B'$.
	The leaf~$w$ they reach lies in~both forced sets; since $L'$ and $L''$ are disjoint, $i_w=i$, so~$x\in S_w$ and $y\in T_w$, and the output~$o_w$ is~valid by~\eqref{eq:global-leaf-output}.

	Finally, Alice can force $L_i=L'\cup\{w:i_w=i,\ x_i\in S_w\}$ exactly when $x_i\in A'$.
	Since $L_{i-1}=L'\cup\{w:i_w=i,\ y_i\notin T_w\}$ is~the complement of~$L''\cup\{w:i_w=i,\ y_i\in T_w\}$, \Cref{lem:global-determinacy} shows that Alice can force $L_{i-1}$ exactly when $y_i\notin B'$.
	Therefore,
	\[k^{-2}\le p_i-p_{i-1}=\Pr[x_i\in A']-\Pr[y_i\notin B']=\mu(A')+\nu(B')-1.\qedhere\]
\end{proof}

\begin{proof}[Proof of~\Cref{thm:global-choice}]
	\Cref{lem:global-weak-separation} gives a~separate rectangle for each pair of~distributions; we~first turn these rectangles into one random rectangle that works for all inputs at~once.
	Let $\mathcal H$ be~the set of~all solvable rectangles; it~is~finite and contains $A\times\varnothing$ and $\varnothing\times B$.
	Recall that $A'\times B'$ covers $x\in A$ if~$x\in A'$, and covers $y\in B$ if~$y\in B'$, and put $\delta=1/2+k^{-2}/4$.
	We~want a~distribution~$\Lambda$ on~$\mathcal H$ such that
	\[\Pr_{H\sim\Lambda}[H\text{ covers }z]\ge\delta\quad\text{for every }z\in A\sqcup B.\]
	By~\Cref{lem:minimax}, applied to~the set~$\mathcal H$, to~$\mathcal Z=A\sqcup B$, and to~the relation $C=\{(H,z):H\text{ covers }z\}$, it~suffices to~show that for every distribution~$\pi$ on~$A\sqcup B$, some $H\in\mathcal H$ satisfies
	\[\Pr_{z\sim\pi}[H\text{ covers }z]\ge\delta.\]

	Fix~$\pi$, and let $\lambda=\Pr_{z\sim\pi}[z\in A]$.
	If~$\lambda\ge\delta$, then $H=A\times\varnothing$ works, since it~covers~$z$ exactly when $z\in A$; similarly, if~$1-\lambda\ge\delta$, then $H=\varnothing\times B$ works.
	Otherwise, $|\lambda-1/2|<k^{-2}/4$.
	Let $\mu$ and~$\nu$ be~the conditional distributions of~$\pi$ on~$A$ and on~$B$, respectively, and let $A'\times B'$ be~the rectangle of~\Cref{lem:global-weak-separation}, so~that $\mu(A')+\nu(B')\ge1+k^{-2}$.
	Then $H=A'\times B'$ works: since $|\mu(A')-\nu(B')|\le1$, we~have 
	\begin{align*}
		\Pr_{z\sim\pi}[H\text{ covers }z]&=\lambda\mu(A')+(1-\lambda)\nu(B') \\
								 &=\frac{\mu(A')+\nu(B')}2+\left(\lambda-\frac12\right)\left(\mu(A')-\nu(B')\right) \\
								 &\ge\frac{1+k^{-2}}2-\frac{k^{-2}}4 \\
								 &= \delta.
	\end{align*}
	Hence there is~a~distribution~$\Lambda$ such that for a~random rectangle $A'\times B'$ from~$\Lambda$,
	\[\Pr[x\in A']\ge\delta\ \text{for every } x\in A,\qquad\Pr[y\in B']\ge\delta\ \text{for every } y\in B.\]

	We~now amplify the probability by~a~majority argument.
	Regard every solvable rectangle $A'\times B'$ as~the function on~$A\sqcup B$ that equals $\mathbf1[x\notin A']$ at~each $x\in A$ and $\mathbf1[y\in B']$ at~each $y\in B$, and let $\mathcal F$ be~the set of~these functions.
	Then $\Lambda$ is~a~distribution over formulas of~depth zero with leaves from~$\mathcal F$, and by~the display above, a~leaf drawn from~$\Lambda$ is~zero at~each $x\in A$ with probability at~least~$\delta$ and one at~each $y\in B$ with probability at~least~$\delta$.
	Since $|A\sqcup B|\le2^{n+1}$, \Cref{lem:majority-amplification} with $\mathcal Z=A\sqcup B$ and $\varepsilon=k^{-2}/4$ gives a~formula~$\Psi$ of~depth $O(\log k+\log n)$ with leaves from~$\mathcal F$ that is~zero at~every $x\in A$ and one at~every $y\in B$.
	On~inputs $x\in A$ and $y\in B$, the players walk down~$\Psi$ from the root while the current subformula is~zero at~$x$ and one at~$y$, as~in~the proof of~\Cref{thm:kw}~\cite{KW90}: at~an~AND gate, both children are one at~$y$, and Alice names a~child that is~zero at~$x$; at~an~OR~gate, both children are zero at~$x$, and Bob names a~child that is~one at~$y$.
	At~the leaf $A'\times B'$ they reach, $x\in A'$ and $y\in B'$, so~they finish by~running a~protocol of~depth at~most $\max\{\CC(\choice^k_{\mathcal R}),\lceil\log_2|O|\rceil\}$ that solves $\mathcal R$ on~$A'\times B'$.
	The resulting protocol for~$\mathcal R$ has depth
	\[\max\{\CC(\choice^k_{\mathcal R}),\lceil\log_2|O|\rceil\}+O(\log k+\log n).\qedhere\]
\end{proof}

Finally, we~derive the bound that holds even for exponentially many instances.

\mainAllArity*

\begin{proof}
	Let $\mathcal R$ be~either $F$ or~$\KW_f$ (with $A=f^{-1}(0)$, $B=f^{-1}(1)$, and $O=[n]$), and let $\Pi$ be~an~optimal protocol for $\choice^k_{\mathcal R}$, of~depth~$d$.
	The leaves of~$\Pi$ name a~set $I\subseteq[k]$ of~at~most $2^d$ positions; put $r=|I|$.
	Given tuples in~$\Inj_r(A)$ and $\Inj_r(B)$, the players place them at~the positions in~$I$, and each player fills the other $k-r$ positions with distinct inputs outside its own tuple by~a~fixed rule, which is~possible since $k\le\min\{|A|,|B|\}$.
	Since $\Pi$ always answers at~a~position in~$I$, running~$\Pi$ on~these tuples solves $\choice^r_{\mathcal R}$ in~depth~$d$.
	Hence \Cref{thm:global-choice} with $r\le2^d$ instances gives
	\[\CC(\mathcal R)\le\max\{d,\lceil\log_2|O|\rceil\}+O(d+\log n),\]
	where $\lceil\log_2|O|\rceil\le1+\log_2n$ in~both cases.
	For $\mathcal R=F$, this is~the first bound; for $\mathcal R=\KW_f$, it~is~the second, since $\CC(\KW_f)=\FD(f)$ by~\Cref{thm:kw} and $d=\FD(\agree^k_f)$ by~\Cref{rem:agreement-kw}.
\end{proof}
 \section{Dense Restrictions and Universal Composition}\label{sec:dense-restrictions}

This section proves our results on~dense restrictions.
For a~nonconstant total $g\colon\bits^n\to\bits$ and a~matrix $X\in\bits^{m\times n}$ with rows $X_1,\ldots,X_m$, write $g^m(X)=(g(X_1),\ldots,g(X_m))$.
We~identify a~matrix with the tuple of~its rows, so~that $S^m\subseteq\bits^{m\times n}$ for every $S\subseteq\bits^n$.
Recall that for a~nonempty subset $\mathcal X\subseteq\bits^{m\times n}$, in~the restriction $\restr{g}{\mathcal X}$ of~$U_m\diamond\KW_g$, the players receive $X,Y\in\mathcal X$ and must output a~cell $(i,j)$ with $X_{i,j}\ne Y_{i,j}$, or~to~reject when $g^m(X)=g^m(Y)$.
We~call any nonempty set $\mathcal X_0\subseteq\bits^{m\times n}$ a~\emph{core}, and the \emph{relative density} of~a~subset $\mathcal X\subseteq\mathcal X_0$ is~$|\mathcal X|/|\mathcal X_0|$.
Unless stated otherwise, the core is~$\bits^{m\times n}$ itself.

\Cref{sec:dense-protocol-encoding} encodes protocols for $\restr{g}{\mathcal X}$ as~formulas with auxiliary variables and turns them into predictors for~$g$.
\Cref{sec:dense-positive-proof} uses these predictors to~prove \Cref{thm:half-density-core}, and \Cref{sec:dense-random-proof} establishes a~bound of~the same form for random functions at~density $2^{1-m}$ in~$\bits^{m\times n}$ in~\Cref{thm:random-dense-core}.
\Cref{sec:dense-composition} derives the composition bounds, and \Cref{sec:dense-negative-proof} proves \Cref{thm:uniform-density-counterexample}.

\subsection{From Protocols to~Formulas}\label{sec:dense-protocol-encoding}

Recall that the KW correspondence~\cite{KW90} turns a~protocol for $\KW_f$ into a~formula for $f$ by~labeling every leaf with a~literal on~which the two inputs differ.
We~apply a~similar idea to~protocols for $\restr{g}{\mathcal X}$.
To~get a~De~Morgan formula, we~need to~decide how to~handle the reject leaves.
Consider a~reject leaf of~a~protocol for $\restr{g}{\mathcal X}$.
Every matrix in~a~pair $(X,Y)$ that reaches this leaf has the same value $c\in\bits^m$ of~$g^m$, that is, $g^m(X)=g^m(Y)=c$.
We~label such a~leaf with an~auxiliary variable~$p_c$.

\begin{lemma}\label{lem:dense-protocol-encoding}
	For every protocol of~depth~$d$ for $\restr{g}{\mathcal X}$ there is~a~De~Morgan formula $\phi(X,p)$ of~depth at~most $d$ in~the matrix variables $X=(X_{i,j})\in\bits^{m\times n}$ and the auxiliary variables $p=(p_c)_{c\in\bits^m}$, such that no~auxiliary variable appears negated and
\[
    \phi(X,p)=p_{g^m(X)}\qquad\text{for every }X\in\mathcal X\text{ and every assignment $p$ to~the auxiliary variables}.
\]
\end{lemma}

\begin{proof}
We~first turn the protocol into a~formula.
Remove all nodes $v$ of~the protocol whose reaching rectangle $\mathcal X_v\times\mathcal Y_v\subseteq\mathcal X\times\mathcal X$ is~empty, and contract every node that has only one remaining child.
At~a~leaf~$v$ that outputs a~cell $(i,j)$, every pair $(X,Y)\in\mathcal X_v\times\mathcal Y_v$ satisfies $X_{i,j}\ne Y_{i,j}$, so~$X_{i,j}$ is~constant on~$\mathcal X_v$ and takes the opposite value on~$\mathcal Y_v$.
We~label this leaf with the literal $X_{i,j}$ or~$\neg X_{i,j}$ that evaluates to~zero on~$\mathcal X_v$ and to~one on~$\mathcal Y_v$.
At~a~reject leaf~$v$, every pair $(X,Y)\in\mathcal X_v\times\mathcal Y_v$ satisfies $g^m(X)=g^m(Y)$, so~all matrices in~$\mathcal X_v\cup\mathcal Y_v$ share a~single value $c\in\bits^m$ of~$g^m$; we~label this leaf with the auxiliary variable~$p_c$.
All reject leaves with the same value~$c$ share the same auxiliary variable~$p_c$.
Replace Alice's nodes by~AND gates and Bob's nodes by~OR gates.

We~now show that this formula satisfies the claimed identity.
For each node $v$, let $\phi_v$ be~the resulting subformula and $\mathcal X_v\times\mathcal Y_v$ its reaching rectangle.
For every assignment~$p$ to~the auxiliary variables, induction gives
\[
    \phi_v(X,p)\le p_{g^m(X)}\quad\text{for } X\in \mathcal X_v
    \qquad\text{and}\qquad
    \phi_v(Y,p)\ge p_{g^m(Y)}\quad\text{for } Y\in \mathcal Y_v.
\]
Indeed, these inequalities hold at~every leaf: a~literal leaf evaluates to~$0\le p_{g^m(X)}$ on~$\mathcal X_v$ and to~$1\ge p_{g^m(Y)}$ on~$\mathcal Y_v$, while a~leaf labeled by~an~auxiliary variable~$p_c$ equals $p_{g^m(X)}$ on~$\mathcal X_v$ and $p_{g^m(Y)}$ on~$\mathcal Y_v$.
At~an~Alice node~$v$, any $X\in \mathcal X_v$ belongs to~$\mathcal X_u$ for one child~$u$, and any $Y\in\mathcal Y_v$ belongs to~$\mathcal Y_u$ for both children, so~taking the AND of~the children's subformulas preserves both inequalities. The case of~a~Bob node is~symmetric.
At~the root~$r$, we~have $\mathcal X_r=\mathcal Y_r=\mathcal X$, so~both inequalities hold for every $X\in\mathcal X$, and hence $\phi(X,p)=p_{g^m(X)}$.
\end{proof}

The next lemma turns this formula into a~formula on~$\bits^n$ that predicts~$g$ under a~distribution~$\nu$ better than a~random guess whenever $\mathcal X$ has density above $2^{-m}$ under~$\nu^m$.

\begin{lemma}\label{lem:dense-grouped-prediction}
Let $m,n\ge1$, let $g\colon\bits^n\to\bits$, let $\mathcal X\subseteq\bits^{m\times n}$ be~nonempty, let $\nu$ be~any distribution on~$\bits^n$, and let $\rho=\Pr_{X\sim\nu^m}[X\in\mathcal X]$ be~the density of~$\mathcal X$ under~$\nu^m$.
If~$\restr{g}{\mathcal X}$ has a~depth-$d$ protocol and $\rho\ge 2^{-m}$, then there is~a~formula $h$ of~depth at~most $d+1$ that satisfies
\[\Pr_{x\sim\nu}[h(x)=g(x)]\ge\frac12+\frac{2^m\rho-1}{2(2^m-1)}.\]
\end{lemma}

\begin{proof}
First, for every set $U\subseteq\bits^m$ we~build a~formula~$\Phi_U$ such that $\Phi_U(X)=\mathbf1[g^m(X)\in U]$ for every $X\in\mathcal X$.
Take the formula $\phi$ from~\Cref{lem:dense-protocol-encoding}.
Since no~auxiliary variable appears negated in~$\phi$, it~is~monotone in~its auxiliary variables.
For a~set $U\subseteq\bits^m$, let $\mathbf1_U\in\bits^{2^m}$ be~the indicator of~$U$.
Consider the formula
\[
    \Phi_U(X)=\phi(X,\mathbf1_U)\wedge\neg\phi(X,\mathbf1_{\bits^m\setminus U}).
\]
On~$\mathcal X$ the second conjunct is~redundant: for $X\in\mathcal X$ we~have $\phi(X,\mathbf1_{\bits^m\setminus U})=1-\phi(X,\mathbf1_U)$, so~$\Phi_U(X)=\phi(X,\mathbf1_U)$.
Outside~$\mathcal X$, \Cref{lem:dense-protocol-encoding} says nothing about $\phi(X,\cdot)$, and it~may happen that $\phi(X,\mathbf1_U)=\phi(X,\mathbf1_V)=1$ for disjoint sets $U$ and~$V$.
The second conjunct rules out $\Phi_U(X)=\Phi_V(X)=1$.
The formula~$\Phi_U$ has depth at~most $d+1$, since negation can be~pushed to~the input literals and constants.
For disjoint $U,V$, monotonicity gives the pointwise inequality
\[
    \Phi_U(X)+\Phi_V(X)\le \Phi_{U\sqcup V}(X).
\]
Indeed, if~either $\Phi_U(X)=1$ or~$\Phi_V(X)=1$, monotonicity in~$p$ implies $\Phi_{U\sqcup V}(X)=1$, and both cannot equal one simultaneously because $U\subseteq\bits^m\setminus V$ implies $\phi(X,\mathbf1_U)\le\phi(X,\mathbf1_{\bits^m\setminus V})$.

Next, we~measure how often $\Phi_U$ is~one when $U$ is~the set of~vectors $c\in\bits^m$ that agree with~$g$ on~the first $i$ rows.
For $X\sim\nu^m$ and $0\le i\le m$, define
\[
    C_i(X)=\{c\in\bits^m:c_j=g(X_j)\text{ for }1\le j\le i\},\qquad M_i
    = \mathbb E_X [ \Phi_{C_i(X)}(X) ].
\]
Here $C_0(X)=\bits^m$ and $M_0\le1$.
Moreover, $C_m(X)=\{g^m(X)\}$, and $\Phi_{\{g^m(X)\}}(X)=1$ for every $X\in\mathcal X$ since $\phi(X,p)=p_{g^m(X)}$. Hence $M_m\ge\rho$.

Let $\alpha$ be~the largest advantage over $1/2$ with which a~formula of~depth at~most $d+1$ predicts~$g$ under~$\nu$, that is,
\[
    \alpha=\max_h\Pr_{x\sim\nu}[h(x)=g(x)]-\frac12,
\]
where $h$ ranges over all formulas of~depth at~most $d+1$ on~$\bits^n$.
We~now show that $M_i\le\frac12M_{i-1}+\alpha$ for every $i\in[m]$.
Informally, $\Phi_U(X)=1$ means that at~$X$ the formula~$\phi$ claims $g^m(X)\in U$, so~$M_i$ is~the probability that $\phi$ claims the correct values of~$g$ on~the first $i$ rows.
Fix $i\in[m]$ and fix all rows of~$X$ except the $i$th row~$X_i$.
The set $C_{i-1}(X)$ depends only on~the first $i-1$ rows, so~it~is~fixed. Partition it~into $U_0=\{c\in C_{i-1}(X):c_i=0\}$ and $U_1=\{c\in C_{i-1}(X):c_i=1\}$.
For $b\in\bits$, let $h_b(x)$ be~the formula in~$x\in\bits^n$ obtained from $\Phi_{U_b}(X)$ by~substituting~$x$ for the $i$th row~$X_i$.
Both $h_0$ and $h_1$ have depth at~most $d+1$.
Thus $h_b(x)=1$ means that, at~the matrix with $i$th row~$x$, the formula $\phi$ claims the correct values on~the first $i-1$ rows and the value~$b$ on~the $i$th row; we~then say for short that $\phi$ claims~$b$ at~$x$.
Since $U_0$ and $U_1$ are disjoint, the pointwise inequality for $\Phi$ above shows that $h_0(x)$ and $h_1(x)$ cannot both equal one, so~$\phi$ claims at~most one value at~$x$.
Hence $h_{g(x)}(x)=1$ exactly when $\phi$ claims the correct value~$g(x)$, and $h_0(x)+h_1(x)=1$ exactly when $\phi$ claims some value.
To~predict $g(x)$, consider a~predictor that outputs the value claimed by~$\phi$ at~$x$, or~a~uniformly random bit if~$\phi$ claims no~value.
This predictor is~the uniform mixture of~the formulas $h_1$ and $\neg h_0$: if~$\phi$ claims a~value, both formulas output this value, and otherwise $h_1$ outputs zero and $\neg h_0$ outputs one.
Hence, for every $x\in\bits^n$, this predictor outputs $g(x)$ with probability
\[
    \frac{\mathbf1[h_1(x)=g(x)]+\mathbf1[\neg h_0(x)=g(x)]}{2}=\frac12+h_{g(x)}(x)-\frac{h_0(x)+h_1(x)}{2}.
\]
Indeed, both sides equal one if~$\phi$ claims $g(x)$, zero if~$\phi$ claims $1-g(x)$, and $1/2$ if~$\phi$ claims no~value.
Both $h_1$ and $\neg h_0$ are formulas of~depth at~most $d+1$, since negation can be~pushed to~the input literals and constants; hence each agrees with~$g$ under~$\nu$ with probability at~most $1/2+\alpha$, and so~does their mixture.
This holds for every fixing of~the rows other than the $i$th.
Since $C_i(X)=U_{g(X_i)}$, we~have $h_{g(X_i)}(X_i)=\Phi_{C_i(X)}(X)$, whose expectation over~$X$ is~$M_i$.
Taking the expectation over $X_i\sim\nu$ and then over the remaining rows of~$X$, and writing $T_i=\mathbb E_X[h_0(X_i)+h_1(X_i)]$ for the probability that $\phi$ claims a~value on~row~$i$, we~obtain $M_i-T_i/2\le \alpha$.
Since $U_0$ and $U_1$ are disjoint with $U_0\sqcup U_1=C_{i-1}(X)$, the pointwise inequality above gives $h_0(X_i)+h_1(X_i)\le \Phi_{C_{i-1}(X)}(X)$, so~$T_i\le M_{i-1}$ and hence
\[
    M_i\le\frac12M_{i-1}+\alpha.
\]

Finally, iterating this inequality for $i=1,\ldots,m$ yields
\[
    \rho\le M_m\le2^{-m}M_0+\alpha\sum_{j=0}^{m-1}2^{-j}\le2^{-m}+2\alpha(1-2^{-m}).
\]
Rearranging gives $\alpha\ge(2^m\rho-1)/(2(2^m-1))$.
Hence a~formula~$h$ attaining the maximum in~the definition of~$\alpha$ satisfies the lemma.
\end{proof}
 
\subsection{Hard Cores for Every Function}\label{sec:dense-positive-proof}

The core $S^m$ of~\Cref{thm:half-density-core} is~built from a~set $S\subseteq\bits^n$ on~which shallow formulas cannot predict $g$ well.
The next lemma shows that such a~set exists whenever $g$ is~hard for formulas.

\begin{lemma}\label{lem:dense-uniform-hard-set}
Let $g\colon\bits^n\to\bits$, and let $t\ge0$ be~an~integer.
Suppose that every nonempty $S\subseteq\bits^n$ admits a~formula of~depth at~most $t$ that agrees with~$g$ on~at~least $2|S|/3$ points of~$S$.
Then $\FD(g)\le t+O(\log n)$.
\end{lemma}

\begin{proof}
We~show that if~$\FD(g)$ is~larger, then rounding a~hard distribution produces a~set $S$ that violates the assumption.
Let $K=\Theta(2^t\log n)$ be~an~integer large enough for the union bound at~the end of~this proof, and let $d=\lceil\log_2n\rceil+\lceil\log_2K\rceil$, so~that $t\le d\le t+O(\log n)$.
A~formula of~depth~$d$ can memorize $g$ on~any $K$ points and output a~fixed constant elsewhere: take a~balanced OR of~the equality tests for the points where $g$ differs from the constant, and complement it~if~necessary.

We~first find a~distribution under which no~formula of~depth at~most~$d$ predicts~$g$ well.
Suppose that $\FD(g)$ exceeds $d$ by~more than the depth that \Cref{lem:majority-amplification} adds for $\varepsilon=1/16$ and $\mathcal Z=\bits^n$, which is~$O(\log n)$.
Then some distribution~$\mu$ on~$\bits^n$ makes every formula~$h$ of~depth at~most~$d$ inaccurate: its accuracy under~$\mu$ satisfies
\[
    \Pr_{x\sim\mu}[h(x)=g(x)]<9/16.
\]
Indeed, otherwise \Cref{lem:minimax}, applied to~the set $\mathcal H$ of~formulas of~depth at~most~$d$, to~$\mathcal Z=\bits^n$, and to~the relation $C=\{(h,x):h(x)=g(x)\}$, gives a~distribution~$\Lambda$ over these formulas such that
\[
    \Pr_{h\sim\Lambda}[h(x)=g(x)]\ge9/16\quad\text{for every }x\in\bits^n,
\]
and \Cref{lem:majority-amplification}, with $\mathcal F$ the set of~literals and constants, turns~$\Lambda$ into a~formula for~$g$ of~depth $d+O(\log n)$, a~contradiction.

Next, we~remove the inputs of~largest probability, since a~formula of~depth~$d$ can memorize~$g$ on~them.
The support of~$\mu$ has more than $K$ points, since otherwise memorization would be~correct everywhere.
Let $T$ be~the set of~the $K$ inputs of~largest probability under~$\mu$.
Memorizing $g$ on~$T$ and using the better constant outside~$T$ gives accuracy at~least $\mu(T)+(1-\mu(T))/2$, so~$\mu(T)<1/8$.
Let $\nu$ be~the conditional distribution of~$x\sim\mu$ given $x\notin T$.
Every input outside~$T$ has probability at~most $\mu(T)/K<1/(8K)$ under~$\mu$, so~$\nu(x) = \mu(x)/(1-\mu(T)) < 1/(7K)$ for every~$x$.
Moreover, every formula of~depth at~most $t\le d$ has accuracy less than $(9/16)/(7/8)=9/14$ under~$\nu$.

Finally, we~round~$\nu$ to~a~set~$S$ that violates the assumption.
Include each point $x$ in~a~random set $S$ independently with probability $K\nu(x)<1$.
For a~fixed formula $h$ of~depth at~most~$t$, let
\[
    w_h=\sum_{x\in\bits^n}\mathbf1[x\in S](\mathbf1[h(x)=g(x)]-2/3)
\]
be~the number of~points of~$S$ on~which $h$ agrees with~$g$, minus $2|S|/3$.
It~is~a~sum of~independent variables bounded by~one in~absolute value, with mean less than $K(9/14-2/3)=-K/42$ and variance at~most~$K$, so~Bernstein's inequality gives $\Pr[w_h\ge0]\le2^{-\Omega(K)}$.
Padding to~a~complete binary tree shows that there are at~most $2^{2^t}(2n+2)^{2^t}=2^{2^t\log_2(4n+4)}$ formulas of~depth at~most~$t$.
Therefore, by~the choice of~$K$, a~union bound over these formulas gives a~set $S$ with $w_h<0$ for every such~$h$.
This set is~nonempty, since $w_h=0$ on~the empty set, and it~violates the assumption.
Hence $\FD(g)\le d+O(\log n)=t+O(\log n)$.
\end{proof}

\mainHalfDensity*

\begin{proof}
We~first choose $S$ as~a~set on~which formulas of~depth about $\FD(g)$ cannot predict~$g$ well.
For an~integer $t\ge0$, call a~nonempty set $S\subseteq\bits^n$ \emph{$t$-hard} if~no~formula of~depth $t$ agrees with~$g$ on~at~least $2|S|/3$ points of~$S$.
In~these terms, \Cref{lem:dense-uniform-hard-set} says that if~no~set is~$t$-hard, then $\FD(g)\le t+O(\log n)$.
If~no~set is~$0$-hard, the lemma with $t=0$ gives $\FD(g)=O(\log n)$, and the theorem holds trivially for every nonempty~$S$.
Otherwise, let $t$ be~the largest integer such that some set is~$t$-hard, and fix a~$t$-hard set~$S$.
This $t$ exists and is~smaller than $\FD(g)$, since a~formula for~$g$ of~depth $\FD(g)$ agrees with~$g$ everywhere, so~no~set is~$t'$-hard for $t'\ge\FD(g)$.
By~the maximality of~$t$, no~set is~$(t+1)$-hard, so~\Cref{lem:dense-uniform-hard-set} with $t+1$ in~place of~$t$ gives $\FD(g)\le t+1+O(\log n)$.

Now let $m\ge2$, let $\mathcal X\subseteq S^m$ have relative density $\rho\ge(1+2^{1-m})/3$, and suppose that $\restr{g}{\mathcal X}$ has a~protocol of~depth~$d$.
Let $\nu$ be~the uniform distribution on~$S$.
Then $\Pr_{X\sim\nu^m}[X\in\mathcal X]=|\mathcal X|/|S|^m=\rho$, and $\rho\ge2^{-m}$, since $(1+2^{1-m})/3\ge2^{-m}$.
Hence \Cref{lem:dense-grouped-prediction} gives a~formula~$h$ of~depth at~most $d+1$ that agrees with~$g$ on~a~fraction at~least
\[
    \frac12+\frac{2^m\rho-1}{2(2^m-1)}\ge\frac12+\frac{(2^m+2)/3-1}{2(2^m-1)}=\frac12+\frac16=\frac23
\]
of~the points of~$S$.
Since $S$ is~$t$-hard, $d+1>t$, that is, $d\ge t$.
This holds for every protocol for $\restr{g}{\mathcal X}$, so~$\CC(\restr{g}{\mathcal X})\ge t\ge\FD(g)-1-O(\log n)$.
\end{proof}

\subsection{Dense Restrictions of~Random Functions}\label{sec:dense-random-proof}

This subsection proves \Cref{thm:random-dense-core}.
We~must show that, for almost every~$g$, no~$\mathcal X\subseteq\bits^{m\times n}$ of~relative density at~least $2^{1-m}$ admits a~shallow protocol for $\restr{g}{\mathcal X}$.
A~union bound over the sets~$\mathcal X$ is~hopeless, since there are $2^{2^{mn}}$ of~them.
Instead, we~use \Cref{lem:dense-protocol-encoding}: a~protocol of~depth~$d$ for $\restr{g}{\mathcal X}$ gives a~formula of~depth at~most~$d$ that by~itself determines $g^m(X)$ for every matrix $X\in\mathcal X$.
Thus a~shallow protocol for such a~set~$\mathcal X$ yields a~shallow formula whose predictor of~$g^m$ is~correct on~at~least a~$2^{1-m}$ fraction of~the matrices.
For a~fixed formula and a~random~$g$, this happens only with tiny probability (see~\Cref{lem:dense-predictor-tail} below), and there are few shallow formulas, so~a~union bound over formulas proves the theorem.

For a~fixed formula $\phi(X,p)$ as~in~\Cref{lem:dense-protocol-encoding} and a~matrix $X\in\bits^{m\times n}$, define $\tau_\phi(X)=c$ if~$\phi(X,p)=p_c$ for every assignment~$p$ to~the auxiliary variables, and put $\tau_\phi(X)=\bot$ otherwise.
Since different auxiliary variables define distinct functions of~$p$, this gives a~well-defined map $\tau_\phi\colon\bits^{m\times n}\to\bits^m\cup\{\bot\}$ that depends only on~$\phi$.
If~$\phi$ is~obtained by~\Cref{lem:dense-protocol-encoding} from a~protocol for~$\restr{g}{\mathcal X}$, then $\phi(X,p)=p_{g^m(X)}$, and hence $\tau_{\phi}(X)=g^m(X)$ for every $X\in\mathcal X$.
However, $\tau_\phi$ is~defined for every De~Morgan formula $\phi(X,p)$ in~the matrix variables~$X$ and the auxiliary variables~$p$, not necessarily obtained from a~protocol for some $\restr{g}{\mathcal X}$, so~the union bound below can range over all such formulas.

Throughout this subsection, put $N=2^n$.
The next lemma shows that a~fixed predictor is~correct on~at~least a~$2^{1-m}$ fraction of~the matrices only with probability $2^{-\Omega(N/m^2)}$ over a~uniformly random function~$g$.

\begin{lemma}\label{lem:dense-predictor-tail}
Let $m\ge1$ satisfy $N\ge4m^2$, and fix a~map $\tau\colon\bits^{m\times n}\to\bits^m\cup\{\bot\}$.
For a~uniformly random function $g\colon\bits^n\to\bits$, put $q_{\tau}(g)=\Pr_{X\in\bits^{m\times n}}[\tau(X)=g^m(X)]$.
Then for every integer $t\ge1$,
\[\mathop{\mathbb E}_g[q_{\tau}(g)^t]\le2^{-mt}\exp\!\left(\frac{m^2t^2}{2N}\right),\qquad\text{and consequently}\qquad \Pr_g[q_{\tau}(g)\ge2^{1-m}]\le2^{-\Omega(N/m^2)}.\]
\end{lemma}

\begin{proof}
We~first bound the $t$th moment of~$q_{\tau}(g)$ in~terms of~the number of~distinct rows of~$t$ random matrices.
Sample $t$ independent uniform matrices in~$\bits^{m\times n}$, and let $J$ be~the set of~distinct rows among their $mt$ rows.
Conditioned on~these $t$ matrices, the event that $\tau(X)=g^m(X)$ for all $t$ matrices requires that none of~them has $\tau(X)=\bot$, that their predicted bits agree whenever two rows coincide, and that $g(x)$ matches the predicted bit at~each of~the $|J|$ distinct rows $x\in J$.
Thus the conditional probability that all $t$ predictions are correct is~either zero or~$2^{-|J|}$, which gives $\mathbb E_g[q_{\tau}(g)^t]\le\mathbb E[2^{-|J|}]$.

Next, we~bound $\mathbb E[2^{-|J|}]$.
Reveal the $mt$ sampled rows one by~one, and let $J_r$ be~the set of~distinct rows among the first $r$ of~them.
Then $r-|J_r|$ is~the number of~the first $r$ rows that repeat an~earlier row.
Conditional on~the revealed rows, the next row repeats one of~them with probability $|J_r|/N$.
Hence
\[\mathbb E[2^{r+1-|J_{r+1}|}\mid\text{first $r$ rows}]=2^{r-|J_r|}\left(\left(1-\tfrac{|J_r|}{N}\right)\cdot1+\tfrac{|J_r|}{N}\cdot2\right)\le2^{r-|J_r|}(1+r/N).\]
Iterating from $J_0=\varnothing$ and using $1+x\le e^x$ gives
\[\mathbb E[2^{-|J|}]\le2^{-mt}\prod_{r=0}^{mt-1}(1+r/N)\le2^{-mt}\exp\!\left(\frac{m^2t^2}{2N}\right).\]

Finally, we~derive the tail bound.
Take $t=\lfloor N\ln2/m^2\rfloor$. Since $N\ln2/m^2\ge2$, we~have $N\ln2/(2m^2)\le t\le N\ln2/m^2$, so~$m^2t^2/(2N)\le(t\ln2)/2$, and Markov's inequality gives
\[\Pr_g[q_{\tau}(g)\ge2^{1-m}]\le\frac{\mathbb E_g[q_{\tau}(g)^t]}{2^{(1-m)t}}\le2^{-t}\exp\!\left(\frac{m^2t^2}{2N}\right)\le2^{-t/2}=2^{-\Omega(N/m^2)}.\qedhere\]
\end{proof}

To~apply this bound to~every protocol at~once, we~count formulas.
For fixed $m$ and~$n$, every leaf of~a~De~Morgan formula $\phi(X,p)$ is~one of~the $2mn$ input literals, the $2^{m+1}$ literals of~the auxiliary variables, or~two constants.
Padding a~formula of~depth at~most $d$ to~a~complete binary tree of~depth~$d$ shows that there are at~most $2^{2^d}(2mn+2^{m+1}+2)^{2^d}=2^{O(2^d(m+\log n))}$ formulas.
Together, \Cref{lem:dense-predictor-tail} and this count give the following bound for all numbers of~rows up~to~about $2^{n/2}$.

\begin{lemma}\label{lem:dense-extended-range}
Let $n\ge4$ and $1\le M\le\sqrt{2^n}/2$.
With probability at~least $1-M2^{-\Omega(2^n/M^2)}$, a~uniformly random function $g\colon\bits^n\to\bits$ satisfies
\[\CC(\restr{g}{\mathcal X})\ge n-3\log_2m-O(\log\log n)\]
simultaneously for every $1\le m\le M$ and every $\mathcal X\subseteq\bits^{m\times n}$ of~relative density at~least $2^{1-m}$.
\end{lemma}

\begin{proof}
Fix $1\le m\le M$ and an~integer $d\ge0$.
Since $m\le M\le\sqrt{2^n}/2$, we~have $N\ge4m^2$, as~\Cref{lem:dense-predictor-tail} requires.
If~some $\mathcal X\subseteq\bits^{m\times n}$ of~relative density at~least $2^{1-m}$ has $\CC(\restr{g}{\mathcal X})\le d$, then by~\Cref{lem:dense-protocol-encoding} some formula $\phi$ of~depth at~most~$d$ has $\tau_{\phi}(X)=g^m(X)$ for every $X\in\mathcal X$, and hence $q_{\tau_{\phi}}(g)\ge2^{1-m}$.
By~\Cref{lem:dense-predictor-tail} and the count above, this happens with probability at~most $2^{O(2^d(m+\log n))-\Omega(N/m^2)}$.
Now let $d$ be~the largest integer for which the first term in~this exponent is~at~most half of~the second, so~that the probability above is~at~most $2^{-\Omega(N/m^2)}$.
If~$d<0$, then $n-2\log_2m-\log_2(m+\log_2n)-O(1)<0$, so~the claimed bound is~negative and the lemma holds trivially for this~$m$ and every~$g$.
Otherwise,
\[d\ge n-2\log_2m-\log_2(m+\log_2n)-O(1)\ge n-3\log_2m-O(\log\log n).\]
Outside this event, every such $\mathcal X$ has $\CC(\restr{g}{\mathcal X})\ge d+1$.
A~union bound over $m\le M$ shows that the bound fails for some~$m$ with probability at~most $\sum_{m\le M}2^{-\Omega(N/m^2)}\le M2^{-\Omega(N/M^2)}$.
\end{proof}

\begin{theorem}[Dense restrictions of~random functions]\label{thm:random-dense-core}
With probability $1-o(1)$, a~uniformly random $g\colon\bits^n\to\bits$ satisfies, for every $1\le m\le n^2$ and every $\mathcal X\subseteq\bits^{m\times n}$ of~relative density at~least $2^{1-m}$,
\[
    \CC(\restr{g}{\mathcal X})\ge n-O(\log n).
\]
\end{theorem}

\begin{proof}
Apply \Cref{lem:dense-extended-range} with $M=n^2$, which satisfies $M\le\sqrt{2^n}/2$ for sufficiently large~$n$.
The conclusion of~the lemma fails with probability at~most $n^22^{-\Omega(2^n/n^4)}=o(1)$, and for every $m\le n^2$ the lower bound is~$n-6\log_2n-O(\log\log n)=n-O(\log n)$.
\end{proof}
 
\subsection{Universal Composition with Random Functions}\label{sec:dense-composition}

This subsection derives the composition bounds from the bounds for dense restrictions by~the strategy of~Dinur and Meir described in~\Cref{sec:introduction}: the protocol is~followed on~diagonal inputs for $m-1$ bits, and a~bound for dense restrictions is~applied to~the remaining square.
The next lemma carries out this strategy.
The walk in~its proof is~the averaging argument behind~\cite[Lemma~9.2]{DM16}.

\begin{lemma}\label{lem:dense-composition-transfer}
Let $g\colon\bits^n\to\bits$ be~nonconstant, let $m\ge1$, and let $d>0$.
If~every $\mathcal X\subseteq\bits^{m\times n}$ of~density at~least $2^{1-m}$ satisfies $\CC(\restr{g}{\mathcal X})\ge d$, then
\[\CC(U_m\diamond\KW_g)\ge m-1+d.\]
\end{lemma}

\begin{proof}
We~find a~node at~depth $m-1$ whose subprotocol solves a~dense restriction.
Starting at~the root with the set of~all diagonal pairs $(X,X)$ for $X\in\bits^{m\times n}$, walk down the protocol tree by~always choosing a~child that contains at~least half of~the remaining diagonal pairs.
After $t$ steps, the set $\mathcal X_t\subseteq\bits^{m\times n}$ of~matrices~$X$ whose diagonal pair $(X,X)$ reaches the current node has density at~least~$2^{-t}$, and the rectangle of~that node contains $\mathcal X_t\times\mathcal X_t$.
For $t\le m-1$, the density of~$\mathcal X_t$ is~at~least $2^{1-m}$, so~the current node cannot be~a~leaf because $\CC(\restr{g}{\mathcal X_t})\ge d>0$.
Hence the walk reaches a~node at~depth~$m-1$, and the subprotocol rooted at~this node solves $\restr{g}{\mathcal X_{m-1}}$ and therefore has depth at~least~$d$, so~the whole protocol has depth at~least $m-1+d$.
\end{proof}

\mainUniversalComposition*

\begin{proof}
By~\Cref{thm:random-dense-core}, with probability $1-o(1)$, simultaneously for every $1\le m\le n^2$, every $\mathcal X\subseteq\bits^{m\times n}$ of~density at~least $2^{1-m}$ satisfies $\CC(\restr{g}{\mathcal X})\ge n-O(\log n)$.
For every such function, \Cref{lem:dense-composition-transfer} gives $\CC(U_m\diamond\KW_g)\ge m+n-O(\log n)$, and the second inequality holds since $\FD(g)\le n+O(\log n)$.
\end{proof}

The bound of~\Cref{cor:random-universal-composition} is~tight up~to~the logarithmic term, since for every~$g$ there is~a~natural protocol solving $U_m\diamond\KW_g$ using $m+\lceil\log_2(m+1)\rceil+\FD(g)$ bits.

For more rows, \Cref{lem:dense-extended-range} gives the following weaker bound.

\begin{corollary}\label{cor:dense-exponential-composition}
Fix $0<\varepsilon<1/3$.
With probability $1-o(1)$, a~uniformly random $g\colon\bits^n\to\bits$ satisfies, simultaneously for every $1\le m\le2^{\varepsilon n}$,
\[\CC(U_m\diamond\KW_g)\ge m+(1-3\varepsilon)n-O(\log\log n).\]
\end{corollary}

\begin{proof}
Take $M=\lfloor2^{\varepsilon n}\rfloor$ in~\Cref{lem:dense-extended-range}.
For sufficiently large~$n$, we~have $M\le\sqrt{2^n}/2$, and the conclusion of~the lemma fails with probability at~most $M2^{-\Omega(2^{(1-2\varepsilon)n})}=o(1)$.
For every $m\le M$, the dense-subset lower bound of~the lemma is~at~least $n-3\varepsilon n-O(\log\log n)$, which is~positive for sufficiently large~$n$ since $\varepsilon<1/3$.
\Cref{lem:dense-composition-transfer} completes the proof.
\end{proof}

\subsection{Limits of~Hard Cores}\label{sec:dense-negative-proof}

This subsection proves \Cref{thm:uniform-density-counterexample}.
We~must find a~function~$g$ such that, for every integer $0\le\sigma\le m-1$, every core $\mathcal X_0\subseteq\bits^{m\times n}$ contains a~subset~$\mathcal X$ of~relative density at~least $2^{-m+\sigma}$ with $\CC(\restr{g}{\mathcal X})\le\FD(g)-m/2+\sigma+O(\log m)$.
Here $\sigma$ measures how much denser than the trivial threshold $2^{-m}$ the subset~$\mathcal X$ must be.
Unlike in~\Cref{thm:half-density-core}, the core~$\mathcal X_0$ is~arbitrary and need not have the form $S^m$ for a~set $S\subseteq\bits^n$.
The function depends only on~$m$, so~it~is~fixed before $\mathcal X_0$ and~$\sigma$ are chosen.
The idea is~that a~protocol for $\restr{g}{\mathcal X}$ does not need a~formula for~$g$: it~suffices to~have a~shallow formula~$\psi$ that agrees with~$g$ on~every row of~every matrix in~$\mathcal X$.
If~moreover only a~few values of~$g^m$ occur in~$\mathcal X$, then Alice can announce $g^m(X)$, Bob can name a~row~$i$ with $g(X_i)\ne g(Y_i)$, and the players can solve the KW game for~$\psi$ on~that row.
Hence we~need a~function of~large formula depth that has shallow predictors correct with high probability at~every input.
\Cref{lem:dense-graph-predictors} constructs such a~function, and \Cref{lem:dense-core-trimming} turns its predictors into a~shallow protocol on~a~dense subset of~any core.

\begin{lemma}\label{lem:dense-graph-predictors}
For every integer $r\ge2$, there is~a~balanced total function $g$ on~$n=2^r+2r+1$ bits with $\FD(g)\ge3r-O(\log r)$.
Moreover, for every integer $q\ge1$, there is~a~distribution over formulas~$h$ of~depth $2r+O(\log r+\log q)$ such that $\Pr_h[h(x)\ne g(x)]\le2^{-q}$ for every $x\in\bits^n$.
\end{lemma}

\begin{proof}
We~first find a~map~$T$ for which checking whether $y=T(u)$ requires large depth.
For an~arbitrary map $T\colon\bits^{2r}\to\bits^{2^r}$, let $f_T(u,y)=\mathbf1[y=T(u)]$ be~the function of~$2r+2^r$ variables that checks whether $y$ is~the value of~$T$ at~$u$.
Different maps give different functions~$f_T$, so~there are $(2^{2^r})^{2^{2r}}=2^{2^{3r}}$ such functions.
Since $2r+2^r\le2^{r+1}$, padding to~a~complete binary tree shows that there are at~most $2^{2^t}(2^{r+2}+2)^{2^t}\le2^{2^t(r+4)}$ formulas of~depth at~most~$t$ on~these variables.
For $t=3r-\lceil\log_2(r+4)\rceil-1$, this is~less than $2^{2^{3r}}$, so~some $f_T$ has depth greater than~$t$, that is, at~least $3r-O(\log r)$. Fix such a~map~$T$.

Next, we~construct the predictors for~$f_T$.
Choose independent uniform $v_1,\ldots,v_q\in\bits^{2^r}$ and define
\[
    H(u,y)=\bigwedge_{j=1}^q\mathbf1[v_j\cdot y=v_j\cdot T(u)],
\]
with dot products over $\mathbb F_2$.
If~$y=T(u)$, then every conjunct is~one, so~$H(u,y)=1=f_T(u,y)$.
If~$y\ne T(u)$, then the $j$th conjunct is~one exactly when $v_j\cdot(y+T(u))=0$.
Since $y+T(u)\ne0$, this happens with probability $1/2$ independently for each~$j$, so~$\Pr[H(u,y)\ne f_T(u,y)]=2^{-q}$.
Every function of~$2r$ bits, in~particular $u\mapsto v_j\cdot T(u)$, is~computed by~a~DNF of~depth $2r+\lceil\log_2(2r)\rceil$, and the parity $v_j\cdot y$ of~at~most $2^r$ bits is~computed by~a~De~Morgan formula of~depth $2\log_22^r=2r$.
Equality adds two levels, and the final conjunction adds $\lceil\log_2q\rceil$, so~$H$ has depth $2r+O(\log r+\log q)$.

Finally, we~add a~balancing bit.
Consider $g(z,u,y)=z\oplus f_T(u,y)$ on~$1+2r+2^r=n$ input variables.
For each $(u,y)$, exactly one value of~$z$ makes $g$ equal to~one, so~$g$ is~balanced.
Substituting $z=0$ into a~formula for~$g$ gives a~formula for~$f_T$ of~no~larger depth, so~$\FD(g)\ge\FD(f_T)\ge3r-O(\log r)$.
The formulas $z\oplus H(u,y)$ have the same error probability as~$H$ and depth larger by~two, which proves the claim.
\end{proof}

The next lemma turns such predictors into a~shallow protocol on~a~dense subset of~any core.

\begin{lemma}\label{lem:dense-core-trimming}
Let $g\colon\bits^n\to\bits$, let $m\ge1$, and let $d\ge0$.
Suppose that there is~a~distribution over formulas~$h$ of~depth at~most~$d$ such that $\Pr_h[h(x)\ne g(x)]\le1/(2m)$ for every $x\in\bits^n$.
For every core $\mathcal X_0\subseteq\bits^{m\times n}$ and every integer $0\le\sigma\le m-1$, some nonempty $\mathcal X\subseteq\mathcal X_0$ has relative density at~least $2^{-m+\sigma}$ in~$\mathcal X_0$ and
\[\CC(\restr{g}{\mathcal X})\le d+\sigma+O(\log(m+1)).\]
\end{lemma}

\begin{proof}
We~first find a~formula that is~correct on~all rows of~many matrices of~the core.
Sample a~formula~$h$ from the distribution.
For each fixed matrix $X\in\mathcal X_0$, a~union bound over its $m$ rows shows that $h(X_i)=g(X_i)$ for all $i\in[m]$ with probability at~least $1-m/(2m)=1/2$ (this does not require the rows to~be~distinct or~independent).
Hence the expected number of~such matrices is~at~least $|\mathcal X_0|/2$, so~we~can fix a~formula~$\psi$ of~depth at~most~$d$ and a~set $\mathcal Y\subseteq\mathcal X_0$ with $|\mathcal Y|\ge|\mathcal X_0|/2$ such that $\psi(X_i)=g(X_i)$ for every $X\in\mathcal Y$ and every $i\in[m]$.

Next, we~keep only the most frequent values of~$g^m$.
Split $\mathcal Y$ according to~the $2^m$ values of~$g^m$, including empty parts.
Let $\mathcal X$ be~the union of~the $K=2^{\sigma+1}\le2^m$ largest parts.
These parts contain at~least a~$K/2^m$ fraction of~$\mathcal Y$, so
\[
    \frac{|\mathcal X|}{|\mathcal X_0|}\ge\frac K{2^m}\cdot\frac{|\mathcal Y|}{|\mathcal X_0|}\ge\frac{2^{\sigma+1}}{2^m}\cdot\frac12=2^{-m+\sigma}.
\]
In~particular, $\mathcal X$ is~nonempty.

Finally, we~give a~protocol for $\restr{g}{\mathcal X}$.
At~most $K$ distinct values of~$g^m$ appear in~$\mathcal X$, so~Alice can send $g^m(X)$ using $\log_2K=\sigma+1$ bits.
Bob then either rejects if~$g^m(X)=g^m(Y)$ or~sends the index of~a~row~$i$ with $g(X_i)\ne g(Y_i)$, using $\lceil\log_2(m+1)\rceil$ bits.
Since $X,Y\in\mathcal X\subseteq\mathcal Y$, we~have $\psi(X_i)=g(X_i)\ne g(Y_i)=\psi(Y_i)$, and both players know which of~these two values is~zero, since they know $g^m(X)$ and~$i$.
Walking down~$\psi$ from the root as~in~the KW correspondence, the players find a~coordinate~$j$ with $X_{i,j}\ne Y_{i,j}$ using at~most $d$~bits and output the cell~$(i,j)$.
In~total, the protocol uses at~most $\sigma+1+\lceil\log_2(m+1)\rceil+d\le d+\sigma+O(\log(m+1))$ bits.
\end{proof}

\mainDenseCounterexample*

\begin{proof}
Set $r=m/2$ and consider the balanced function $g_m$ from~\Cref{lem:dense-graph-predictors}, so~that $n=2^{m/2}+m+1$ and $\FD(g_m)\ge3m/2-O(\log m)$.
With $q=\lceil\log_2(2m)\rceil$, its predictors have error at~most $1/(2m)$ and depth $m+O(\log m)$.
\Cref{lem:dense-core-trimming} gives a~set $\mathcal X$ of~relative density at~least $2^{-m+\sigma}$ in~$\mathcal X_0$ with
\[\CC(\restr{g_m}{\mathcal X})\le m+\sigma+O(\log m)\le\FD(g_m)-m/2+\sigma+O(\log m).\qedhere\]
\end{proof}

In~particular, let $s(m),\ell(m)\ge0$ be~any functions that are~$o(m)$.
Taking $\sigma=\lceil s(m)\rceil$ in~\Cref{thm:uniform-density-counterexample} shows that, for all sufficiently large even~$m$, no~core $\mathcal X_0$ guarantees $\CC(\restr{g_m}{\mathcal X})\ge\FD(g_m)-\ell(m)$ for all $\mathcal X\subseteq\mathcal X_0$ of~relative density at~least $2^{-m+s(m)}$, because $2^{-m+\lceil s(m)\rceil}\ge2^{-m+s(m)}$ and $m/2-\lceil s(m)\rceil-O(\log m)>\ell(m)$.

\paragraph{Tightness of~\Cref{thm:half-density-core}.}
The same construction shows that the logarithmic loss in~\Cref{thm:half-density-core} is~necessary.
Let $r\ge2$ and $m\ge1$ be~integers, take $q=\lceil\log_2(2m)\rceil$ in~\Cref{lem:dense-graph-predictors}, and apply \Cref{lem:dense-core-trimming} with $\sigma=m-1$.
For the resulting function~$g$ on~$n=2^r+2r+1$ bits, every core $\mathcal X_0\subseteq\bits^{m\times n}$ contains a~subset $\mathcal X$ of~relative density at~least $1/2$ with
\[
    \FD(g)-\CC(\restr{g}{\mathcal X})\ge r-m-O(\log r+\log m).
\]
Adding dummy variables changes neither $\FD(g)$ nor the predictors.
Hence, for every $n\ge9$, we~may take the largest~$r$ with $2^r+2r+1\le n$, which satisfies $n<2^{r+1}+2(r+1)+1\le2^{r+2}$, that is, $r>\log_2n-2$.
This gives a~function on~$n$ bits whose loss at~density~$1/2$ is~at~least $\log_2n-m-O(\log\log n+\log m)$.
For $m\le(\log_2n)/2$, this loss is~$\Omega(\log n)$ and exceeds $0.99m$ for all sufficiently large~$n$.
Since $(1+2^{1-m})/3\le1/2$ for $m\ge2$, this applies in~particular to~the core $S^m$ of~\Cref{thm:half-density-core}, so~its $O(\log n)$ loss is~tight up~to~a~constant factor.

This also answers the question of~Dinur and Meir~\cite{DM16} whether every~$g$ has a~core in~which $\CC(\restr{g}{\mathcal X})\ge\FD(g)-0.99m$ for every subset~$\mathcal X$ of~relative density~$1/2$.
When $m$ is~at~least a~large constant times $\log n$, the answer is~positive by~\Cref{thm:half-density-core}.
When $m\le(\log_2n)/2$ and $n$ is~sufficiently large, the answer is~negative, since the bound fails in~every core for the function~$g$ above.
 \section{Open Problems}\label{sec:scope}

We~conclude with the questions that we~find most interesting.

\paragraph{Direct sum.}
For a~total $F\colon\bits^n\times\bits^n\to\bits$ and $k\ge1$, let $F^k$ be~the function that receives $k$ independent input pairs and outputs all $k$ values of~$F$.

\begin{question}\label{ques:direct-sum}
	Is~there an~absolute constant $c>0$ such that
	\[\CC(F^k)\ge c\,k\,\CC(F)\]
	for every total~$F$ and every~$k$?
\end{question}

For arbitrary total~$F$, the amortized bounds of~Feder, Kushilevitz, Naor, and Nisan~\cite{FKNN95} lose a~square root.
Mackenzie and Saffidine~\cite{MS25} refuted the strongest form $\CC(F^k)\ge k(\CC(F)-O(1))$, but their counterexample does not rule out $\CC(F^k)\ge c\,k\,\CC(F)$.

\paragraph{Choice and agreement for exponentially many instances.}
Our bounds lose $O(\log k)$, so~the question remains open for exponentially many instances.

\begin{question}\label{ques:arbitrary-arity}
	Is~it~true that
	\[\CC(\choice^k_F)\ge\CC(F)-\polylog(n)\qquad\text{and}\qquad\FD(\agree^k_f)\ge\FD(f)-\polylog(n)\]
	for every total $F\colon\bits^n\times\bits^n\to\bits$ with $1\le k\le2^n$ and every nonconstant total $f\colon\bits^n\to\bits$ with $1\le k\le\min\{|f^{-1}(0)|,|f^{-1}(1)|\}$, respectively?
\end{question}

These are Conjectures~9.7 and~9.6 of~Dinur and Meir~\cite{DM16}; the agreement question for partial functions is~even stronger.
\Cref{thm:global-boolean,thm:global-agreement} settle both conjectures for $k\le2^{\polylog(n)}$, and \Cref{cor:global-all-arity} preserves a~constant fraction of~the complexity even for exponentially many instances, up~to~an~additive $O(\log n)$ term.

\paragraph{A~corrected dense-core conjecture.}
\Cref{thm:uniform-density-counterexample} shows that the uniform formulation of~Conjecture~9.4 is~false, which raises the question of~how to~repair~it.

\begin{question}\label{ques:dense-core}
Does an~explicit $O(\log n)$ loss suffice to~repair the uniform formulation of~Conjecture~9.4?
More precisely, are~there absolute constants $C,c>0$ such that, for every $m,n$ and nonconstant $g$, some nonempty core $\mathcal X_0$ satisfies
\[\frac{|\mathcal X|}{|\mathcal X_0|}\ge2^{-m+C\sqrt m(\log(m+2))^c}\quad\Longrightarrow\quad\CC(\restr{g}{\mathcal X})\ge\FD(g)-C\left(\sqrt m(\log(m+2))^c+\log n\right)\]
for every $\mathcal X\subseteq\mathcal X_0$ whenever the density threshold is~at~most one?
\end{question}

\Cref{thm:random-dense-core} establishes this statement for almost every~$g$: it~gives a~stronger bound for $m\le n^2$, and the requested lower bound is~nonpositive for $m>n^2$.
The unresolved case is~an~arbitrary inner function.
\Cref{thm:half-density-core} gives hard cores for every~$g$ at~constant relative density, while \Cref{thm:uniform-density-counterexample} shows why the dependence on~$n$ cannot simply be~omitted.

\paragraph{From choice to~ordinary composition.}
Our composition results concern the universal outer relation, and they do~not give the KRW conjecture for Boolean outer functions.

\begin{question}\label{ques:ordinary-composition}
How can lower bounds for the 1-out-of-$k$ KW game or~for dense restrictions be~used to~prove lower bounds for ordinary composition $f\diamond g$?
\end{question}

In~$\KW_{f\diamond g}$, the players may output a~differing cell in~a~row where the two inner values $g(X_i)$ and $g(Y_i)$ are equal, whereas in~the 1-out-of-$k$ game of~$\KW_g$ every pair of~corresponding rows has opposite $g$-values.
Moreover, the proof of~\Cref{cor:random-universal-composition} for $U_m\diamond\KW_g$ tracks the protocol on~diagonal pairs $(X,X)$, which do~not appear in~$\KW_{f\diamond g}$ because Alice's and Bob's inputs have opposite outer values $f(g^m(A))\ne f(g^m(B))$.
 \section*{AI Methodology Statement}

We~follow the recommendations of~the working group report~\cite{AITCS26} and~use its template for~this section.

\begin{flushleft}
	\renewcommand{\arraystretch}{1.3}
	\begin{tabular}{@{}l@{\qquad}c@{}}
		\textbf{Activity} & \textbf{AI used} \\
		Asking the research question & No \\
		Coming up with the approach & Yes \\
		Proof development & Yes \\
		Writing and exposition & Yes \\
		Checking for bugs & Yes \\
		Other supporting tasks & Yes \\
	\end{tabular}
\end{flushleft}

\paragraph{Details (elaborating on~the points above).}
We~used OpenAI ChatGPT and Codex to~develop candidate proofs, check sources, and draft the text. We used Anthropic Claude Opus 5.5 and Google Antigravity to~restructure the paper and rewrite its exposition, including the abstract and the introduction.
The authors are responsible for all statements.
 
\clearpage
\bibliographystyle{alpha}
\newcommand{\etalchar}[1]{$^{#1}$}

\end{document}